\newif\ifuseprd
\newif\ifom
\useprdfalse 
\omtrue      
\ifuseprd 
\documentclass[preprint,tightenlines,floats,prd,eqsecnum,nobibnotes,%
nofootinbib]{revtex4}
\else
\documentclass[final,12pt,letterpaper]{JHEP}
\fi 
\usepackage{amsmath,amssymb}
\usepackage[final]{epsfig}

\allowdisplaybreaks[1]
\ifom
\font\smalldn=dvng8
\font\ninedn=dvng9
\font\dvng=dvng10
\font\halfdn=dvng10 scaled\magstephalf
\font\bigdn=dvng10 scaled\magstep1
\font\largedn=dvng10 scaled\magstep2
\font\hugedn=dvng10 scaled\magstep3
\def\sethyph#1{
\hyphenchar\smalldn=#1
\hyphenchar\ninedn=#1
\hyphenchar\dvng=#1
\hyphenchar\halfdn=#1
\hyphenchar\bigdn=#1
\hyphenchar\largedn=#1
\hyphenchar\hugedn=#1}
\font\smallcr=cmr8
\font\ninecr=cmr9
\font\halfcr=cmr10 scaled\magstephalf
\font\bigcr=cmr10 scaled\magstep1
\font\largecr=cmr10 scaled\magstep2
\font\hugecr=cmr10 scaled\magstep3
\let\rsize=\rm
\newcount\chnum
\newdimen\itdim
\newdimen\dnblskip
\newif\ifdnmode
\def\subscr#1{\/\itdim=\lastkern
\unkern\kern-\itdim \lower\dp0 \hbox to\itdim{#1\hfil}}
\def\dnsmall{\let\pdn=\smalldn\let\rsize=\smallcr%
\dnblskip=12pt\ifdnmode\dn\fi}
\def\dnnine{\let\pdn=\ninedn\let\rsize=\ninecr%
\dnblskip=13pt\ifdnmode\dn\fi}
\def\dnnormal{\let\pdn=\dvng\let\rsize=\rm%
\dnblskip=15pt\ifdnmode\dn\fi}
\def\dnhalf{\let\pdn=\halfdn\let\rsize=\halfcr%
\dnblskip=16pt\ifdnmode\dn\fi}
\def\dnbig{\let\pdn=\bigdn\let\rsize=\bigcr%
\dnblskip=18pt\ifdnmode\dn\fi}
\def\dnlarge{\let\pdn=\largedn\let\rsize=\largecr%
\dnblskip=22pt\ifdnmode\dn\fi}
\def\dnhuge{\let\pdn=\hugedn\let\rsize=\hugecr%
\dnblskip=26pt\ifdnmode\dn\fi}
\def\dn{\dnmodetrue\pdn\baselineskip=\dnblskip
\chnum=0
\loop\catcode\chnum=11
\ifnum\chnum<12\advance\chnum by1
\repeat
\chnum=14
\loop\catcode\chnum=11
\ifnum\chnum<31\advance\chnum by1
\repeat
\catcode127=11
\tolerance=10000
\pretolerance=10000}
\def\0{\llap{\char13}}
\def\1{\llap{\char32}}
\def\2{\llap{\char92}}
\def\3#1w{{\char"#1}}
\def\4{\llap{\char123}}
\def\5{\llap{\char125}}
\def\6#1{\setbox0=\hbox{#1}#1\subscr{\char126}}
\def\7#1{\setbox0=\hbox{#1}#1\subscr{\char0}}
\def\8#1{\setbox0=\hbox{#1}#1\subscr{\char1}}
\def\9#1{\setbox0=\hbox{#1}#1\subscr{\char2}}
\def\qa#1#2{\setbox0=\hbox{#1}#1\subscr{\char253\kern1.5ex\lower1.25ex
\hbox{\char#2}\kern-1.5ex}}

\def\qva{\kern0.5ex\2\kern-0.5ex}
\def\qvb{\kern1ex\0\kern-1ex}
\def\qvc{\kern1ex\rdt\kern-1ex}
\def\?{\llap{\char3}}
\def\<{\llap{\char4}}

\def\rdt{\llap{\char19}}
\def\dnnum{\let\nstyle=d}
\def\cmnum{\let\nstyle=r}
\cmnum
\def\rn#1{\if\nstyle r{\rsize #1}\else#1\fi}

\sethyph{-1}
\let\pdn=\dvng
\fi 

\def\omt{{\ifom{{\dn\dnhalf :}}\else%
        {{3\!{\footnotesize$\mathbf{{\frown}\llap{\text{\tiny$\prime$}}}$}%
        {\hbox to -.7ex{\null}\llap{\raise1.3ex\hbox{\tiny{%
        \setbox255=\hbox{$\mathbf{{\smile}}$}%
        \copy255\kern-.7\wd255{\raise.5ex\hbox{$\mathbf{\cdot}$}}}}}}}}\fi}}

\newcommand\skipthis[1]{{}}

\newcommand\ct[1]{{\ifuseprd{\em{#1}},\else{\sf {#1}},\fi}}
\newcommand\bt[1]{{\em {#1}},}
\newcommand\web[1]{{\tt \hbox{{#1}}}}
\newcommand\phepth[1]{{\tt [\hepth{#1}]}}

\DeclareMathOperator{\im}{Im}
\DeclareMathOperator{\real}{Re}

\DeclareMathOperator{\Log}{Log}

\newenvironment{spmatrix}{\left(\begin{smallmatrix}}{\end{smallmatrix}\right)}

\chardef\til=`~

\newcommand\p{\ensuremath{\partial}}
\newcommand\evalat[2]{\ensuremath{\left.{#1}\right|_{#2}}}
\newcommand\abs[1]{\ensuremath{\left\lvert{#1}\right\rvert}}
\newcommand\no[1]{{{:}{#1}{:}}}

\newcommand\field[1]{{\ensuremath{\mathbb{{#1}}}}}
\newcommand\order[1]{{\ensuremath{{\mathcal O}({#1})}}}
\newcommand\vev[1]{{\ensuremath{\left\langle{#1}\right\rangle}}}

\DeclareMathOperator{\Tr}{Tr}

\newcommand\lvec[2][]{\ensuremath{\overleftarrow{{#2}_{#1}}}}
\newcommand\rvec[2][]{\ensuremath{\overrightarrow{{#2}_{#1}}}}

\newcommand\uhp{{\ensuremath{{\mathcal{H}}^+}}}

\newcommand\apr{{\ensuremath{{\alpha'}}}}
\newcommand\g{{\ensuremath{{\mathcal G}}}}

\newcommand\ep{\epsilon}

\newcommand\vt{\vartheta}

\providecommand\FIGURE[2][]{\begin{figure}[#1]\begin{center}{#2}\end{center}
                       \end{figure}}
\providecommand\putabstract[1]{\ifuseprd\begin{abstract} {#1} \end{abstract}%
                           \else \abstract{{#1}} \fi}
\providecommand\plb[3]{{Phys.\ Lett.\ B {\bf {#1}}, {#3} ({#2})}}
\providecommand\npb[3]{{Nucl.\ Phys.\ {\bf B{#1}}, {#3} ({#2})}}
\providecommand\jhep[3]{{J.\ High Energy Phys.\ {\bf #1}, {#3} ({#2})}}
\providecommand\npps[3]{{Nucl.\ Phys.\ {bf {#1}} Proc.\ Suppl.\ {#3} ({#2})}}
\providecommand\lmp[3]{{\ifuseprd\else\begingroup\em\fi Lett.\ Math.\ Phys.\ %
     \ifuseprd\else\endgroup\fi {\bf {#1}}\ifuseprd, {#3} ({#2})\else%
     \ ({#2}) {#3}\fi}}
\providecommand\jdf[3]{{\ifuseprd\else\begingroup\em\fi J.\ Diff.\ Geom.\ %
     \ifuseprd\else\endgroup\fi {\bf {#1}}\ifuseprd, {#3} ({#2})\else%
     \ ({#2}) {#3}\fi}}
\newcommand\citeprd[3]{{\ifuseprd{Phys.\ Rev.\ D {\bf {#1}}, {#3} ({#2})}%
                        \else{\prd{#1}{#2}{#3}}\fi}}
\providecommand\hepth[1]{{\tt hep-th/{#1}}}
\newenvironment{smaleq}{\ifuseprd\else\small\fi}{}

\ifuseprd
\begin{document} 
\fi 

\title{\LARGE $\ast$-Trek: \\
\Large The One-Loop ${\mathcal N}=4$ Noncommutative SYM Action}
\ifuseprd
\author{Hong Liu}\email{liu@physics.rutgers.edu}
\author{Jeremy Michelson}\email{jeremy@physics.rutgers.edu}
\affiliation{New High Energy Theory Center \\ 
       Rutgers University \\
       126 Frelinghuysen Road \\
       Piscataway, NJ \ 08854}
\else 
\author{Hong Liu\thanks{\tt liu@physics.rutgers.edu} \ and Jeremy
Michelson\thanks{\tt jeremy@physics.rutgers.edu} \\
New High Energy Theory Center \\
Rutgers University \\
126 Frelinghuysen Road \\
Piscataway, NJ \ 08854 \ USA}
\fi 

\putabstract{
We investigate
${\mathcal N} =4$ noncommutative super Yang-Mills (SYM) theory.  
We compute the one-loop four gauge boson scattering amplitude on 
parallel D$p$-branes, and find the corresponding contribution to the
noncommutative SYM one-loop action in a momentum expansion.  
The result is somewhat surprising. We find that while the planar diagram can 
be written using the usual $\ast$-product, the contributions 
from nonplanar diagrams in general involve additional structure beyond 
the $\ast$-product, arising from the nontrivial worldsheet correlations 
surviving the field theory limit. To each nonplanar diagram, depending on 
the number $n$ of external vertex operator insertions on each boundary, 
there is a corresponding $\ast_n$ $n$-ary operation. We further find
that it is no 
longer possible to write down an off-shell gauge invariant one-loop 
effective action using the noncommutative field strength defined 
at tree-level.
}

\preprint{RUNHETC-2000-31\ifuseprd,~\else\\\fi {\tt hep-th/0008205}}

\ifuseprd
\maketitle
\else
\begin{document}
\fi 

\section{Introduction} \label{sec:intro}

Field and string theories on noncommutative space(time) have attracted
much attention recently~\cite{cds,dh}. They appear naturally in various  
decoupling limits of the worldvolume theories of D-branes in a background 
NS-NS $B$-field~\cite{sw,gmms,sst,gmss,bb,km}. 
While much progress has been made in understanding the perturbative 
dynamics~\cite{filk,chep,mrs,rs} 
and their strong coupling limits~\cite{gmms,sst,gs,gmss,km}, many fundamental 
issues remain obscure.

In noncommutative gauge theories, gauge invariance becomes much more subtle. 
The gauge group and the allowed representations of the gauge group 
are highly constrained. A basic example is that 
in noncommutative field theory, $SO(2)$ is not a good gauge group even
though the isomorphic $U(1)$ is.  The first statement follows from the
fact that
\begin{multline}
\begin{spmatrix} 0 & f(x) \\ -f(x) & 0 \end{spmatrix}
\ast \begin{spmatrix} 0 & g(x) \\ -g(x) & 0 \end{spmatrix} 
- \begin{spmatrix} 0 & g(x) \\ -g(x) & 0 \end{spmatrix} 
\ast \begin{spmatrix} 0 & f(x) \\ -f(x) & 0 \end{spmatrix} \\ 
= \begin{spmatrix} -f(x)\ast g(x) + g(x)\ast f(x) & 0 \\ 0 & -f(x)\ast
g(x) + g(x)\ast f(x)  \end{spmatrix},
\end{multline}
which is not an element of the $SO(2)$ algebra. 
It is also interesting that, in theories with only adjoint matter,
translations along noncommutative directions are a subset of gauge
transformations, and thus there are
no gauge invariant local operators~\cite{gross,dr}.%
\footnote{Matter in other representations can
sometimes be used to construct gauge invariant local operators.\cite{rr}}

Another puzzling issue is the contributions of the nonplanar diagrams
to the 1PI effective action of the noncommutative field
theories. One generically encounters IR singularities 
when an external momentum crosses an internal line in a nonplanar
diagram for which the
internal momentum integration is UV divergent at $\Theta =0$;
this is the so-called IR/UV mixing. 
The presence of IR/UV mixing indicates that the
long distance behaviour of the system is no longer insensitive to the
short distance physics, and the UV cutoff might not be pushed to
infinity in a consistent way (see, however,~\cite{gubser}).

\FIGURE{
\includegraphics[width=2in]{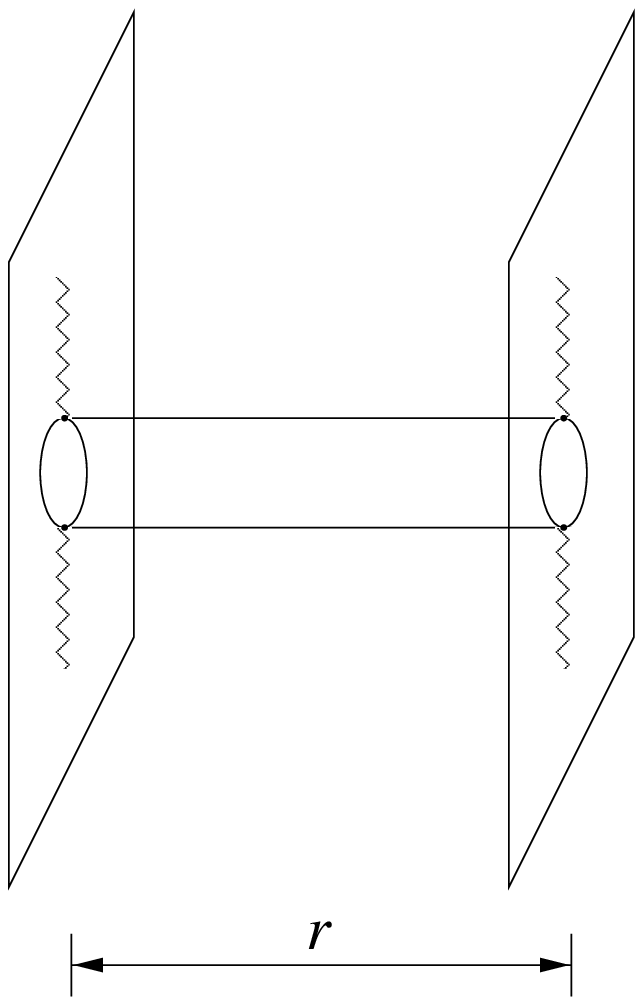}
\caption{An example of the scattering amplitude considered.  This is
the nonplanar diagram with two gluons on each brane (a $2\rightarrow2$
process). \label{fig:bw}}
}

In~\cite{lm}, it was suggested that IR/UV mixing may be naturally understood 
in terms of a ``stretched string''. When $B \neq 0$, an open string
has a nonzero stretched displacement $\Delta x = \Theta k$,%
\footnote{$k$ is the center of mass momentum of the string. 
Semi-classically, the length $\Delta x$ may
be understood via the dipole picture discussed in~\cite{bs,yin,sj}.}
even in the field theory limit $\apr \rightarrow 0$. As  
the momentum $k$ is increased, the string stretches longer.  
Thus when there is a 
net momentum flow $k$ between the two ends of the open string, as in the case 
of a nonplanar diagram, the integrated loop particle has to 
propagate  an additional distance $\abs{\Delta x}$, 
the effect of which 
is to seemingly cut open an otherwise closed first-quantized particle loop.
When, for \hbox{$B=0$}, the loop
integral is divergent at short distances, 
the stretched string acts as an effective short-distance 
cutoff and regularizes the divergence. However, since taking
$\Delta x$ to zero (IR for external momentum $k$) will recover the original 
UV divergence, the amplitude
is generally singular as $\Delta x \rightarrow 0$. Thus,
$k \rightarrow 0$ is not a continuous limit, and the singularity at small $k$
reflects the UV divergence of the $B=0$ theory. 
The stretched string can also be used to understand two-loop
computations.\cite{klp}

In this paper we further explore the IR/UV effects of the stretched 
string. To keep the IR singularities from obscuring our understanding,
we consider a highly supersymmetric example: ${\mathcal N} = 4$ 
noncommutative SYM
theory. Since this theory is finite in the commutative limit, the nonplanar
diagrams are not plagued by the IR singularities mentioned above. We
are thus able to give an explicit and unambiguous result for its
one-loop scattering amplitudes. Various issues, including the
contributions from nonplanar diagrams and one-loop off-shell gauge
invariance can then be discussed in a controlled manner.

Like the commutative theory, there is no 1-loop contribution to the 
two and three-point amplitudes in ${\mathcal N}=4$ noncommutative SYM theory. 
We will thus focus on the four-point amplitudes, 
specifically the four gauge
boson annulus amplitude ({\em e.g.\ \/}figure~\ref{fig:bw}). The leading
order contribution in the low energy limit determines the $F^4$
term in the low energy effective action of noncommutative ${\mathcal N} =4$
super-Yang-Mills theory. Since we are interested in exploring the effect of
noncommutativity, we take the low energy limit to mean%
\footnote{The open string metric $G_{\mu\nu}$ and noncommutativity
parameter $\Theta^{\mu\nu}$ are related to the closed string metric
$g_{\mu\nu}$ and the $B$ field by
\hbox{$G^{\mu\nu}=\bigl(\frac{1}{g+B}g\frac{1}{g-B}\bigr)^{\mu\nu}$}
and
\hbox{$\Theta^{\mu\nu}=-{2\pi\apr}
\bigl(\frac{1}{g+B}B\frac{1}{g-B}\bigr)^{\mu\nu}$}.}
\begin{equation} \label{kin}
\frac{k_i G^{-1} k_j}{m^2} \ll 1  \qquad  \text{with} \qquad
k_i \Theta k_j  \sim \order{1},
\end{equation}
for external momenta $k_i$, where $m$ is the mass of the W-bosons stretched
between the branes.

Before we begin, let us recall the commutative result.  For
convenience, we restrict ourselves to the case of two single branes. 
To lowest order in external momenta, the one-loop amplitude for 
four gauge bosons 
can be written in the 
form (see {\em e.g.\/}~\cite{dougt,das})
\begin{equation} \label{kthree}
\Gamma_{\rm {1-loop}}
\propto \frac{1}{r^4} \int d^4y \bigl[ F^\mu{_\nu}
F^\nu{_\rho}  F^\rho{_\kappa}  F^\kappa{_\mu}  - \frac{1}{4}
F^\mu{_\nu}  F^\nu{_\mu}  F^\rho{_\kappa} F^\kappa{_\rho}\bigr].
\end{equation}
where 
$F=F_1-F_2$ and 
the subscripts $1,2$ refer to the two D3-branes. 

The factor of $\frac{1}{r^4}$ in equations~\eqref{kthree} 
arises in the SYM box diagram via the $W$-boson mass [$m=\frac{r}{2\pi\apr}$]
in the propagators.  Alternatively, one can understand it
as arising via the propagator for exchange of massless
closed strings in the transverse space between the 
branes.

When we turn on a nonvanishing $\Theta$, 
we find some new interesting phenomena for the nonplanar amplitudes:
\begin{itemize}
\item Due to the extra propagating distance $\abs{\Delta x} = |\Theta k|$ 
for the W-bosons running in the loop, where $k$ now is the momentum flow 
between the branes, 
the factor $\frac{1}{r^4}$ is replaced by
\begin{equation} \label{bessel}
\frac{1}{m^4} K_2 (m \abs{\Delta x}), \qquad m=\frac{r}{2\pi\apr}.
\end{equation}
where $K_2$ is a  modified Bessel function. 
In particular, when $m \abs{\Delta x} \gg 1$, the amplitude is exponentially 
suppressed, proportional to
\begin{equation}
\exp [ - m \abs{\Delta x}]
\end{equation}
typical of an intermediate particle of mass $m$ propagating
over a distance $\abs{\Delta x}$.
From the closed string point of view, equation~\eqref{bessel} implies
that the closed strings propagating between the
branes are massive. Due to 
different scalings between the closed and open string metrics, 
the low energy limit on the brane in terms of open string 
metric $G_{\mu \nu}$ no longer implies 
the low energy limit for the bulk closed string modes. 
Thus, in the kinematic regime~\eqref{kin}, there is a significant longitudinal
momentum transfer between the branes, 
and the closed string modes indeed obtain an effective mass 
$M = \frac{\abs{\Delta x}}{2 \pi \apr} =
\frac{|\Theta k|}{2 \pi \apr}$.
\item We have attempted to write down a one-loop 
effective action, from integrating out massive W-bosons,
in terms of a derivative expansion in the regime~\eqref{kin}. We find
that while it is possible to formally write down
an off-shell effective action  from the gauge invariant 
on-shell amplitudes, the off-shell action is not gauge invariant in terms of 
the noncommutative $F_{\mu \nu}$,
\begin{equation}
F_{\mu \nu} = \partial_{\mu} A_{\nu} 
- \partial_{\nu} A_{\mu} - i A_{\mu} \ast A_{\nu} + i A_{\nu} \ast A_{\mu},
\end{equation}
defined at tree-level. The technical reason is the following.
In nonplanar processes, writing an effective action
requires additional 
$\ast$-operations
which are not gauge invariant under the gauge 
transformations of the noncommutative $F$.%
\footnote{For the full expression
for the effective action, see equation~\eqref{ncf4}.}
For example, for the $2 \rightarrow 2$ process of figure~\ref{fig:bw},
the product
between two photon modes on the same worldsheet 
boundary becomes
\begin{equation} 
F^{\mu\nu} \ast' F_{\mu\nu},
\end{equation}
where the
$\ast'$-product is given by
\begin{equation} \label{def*'}
\begin{split}
f(x) \ast' g(x) &\equiv f(x) 
\frac{\sin\bigl(\frac{1}{2}\Theta^{\mu\nu}\lvec[\mu]{\p}\rvec[\nu]{\p}\bigr)}%
{\frac{1}{2}\Theta^{\mu\nu}\lvec[\mu]{\p}\rvec[\nu]{\p}}
g(x), \\ &= f(x) g(x) - \frac{1}{24} \Theta^{\mu\nu} \Theta^{\sigma\tau}
\p_\mu \p_\sigma f(x) \p_\nu \p_\tau g(x) + \ldots \ .
\end{split}
\end{equation}
The result is somewhat surprising. From the field theory point of view,
we have integrated out the massive $W$-bosons. According 
to conventional wisdom, we should be able to write down a gauge-invariant 
effective action as a derivative expansion. That it is no longer 
possible to do so might be another indication of IR/UV mixing in 
noncommutative  theories, even though in this case we do not have the 
IR singularities.

\end{itemize}

The paper is organized as follows.  In section~\ref{sec:calc}, we
evaluate the string amplitude.  The field theory limit is explored in
section~\ref{sec:ft}.  
In section~\ref{sec:conc}, we conclude with a discussion of  gauge
invariance.
We
describe our treatment of the fermionic string coordinates in the
presence of a $B$-field, used in
section~\ref{sec:calc}, in appendix~\ref{sec:fermions}.

\section{Correlation Functions} \label{sec:calc}

We are interested in computing the one-loop scattering amplitudes of 
``brane waves''  on parallel D$p$-branes in type II string theory---%
specifically,  the four gauge boson
annulus amplitude, both planar and nonplanar ({\em e.g.\/}\
figure~\ref{fig:bw}). 

On the annulus, we can work in the ``zero''
picture, for which the relevant vertex operator is~\cite{jp}
\begin{equation} \label{aop}
{\mathcal V}^0 = \frac{g_o}{\sqrt{2 \apr}} T^a e_\mu^a (i \Dot{X}^\mu +
4 k \cdot \Psi \Psi^\mu) e^{i k \cdot X}.
\end{equation}
In equation~\eqref{aop}, 
$T^a$ is a generator of the gauge group,
$e^a_\mu$ is the polarization of the gluon
and $\Psi^\mu$ is the linear
combination of the left and right-moving fermions, $\psi$ and
$\Tilde{\psi}$, that lives on the boundary (see
appendix~\ref{sec:fermions}).
The dot product is simple contraction: $k\cdot\Psi \equiv k_\mu \Psi^\mu$.
For general periodicity \hbox{$\psi^\mu(w+2n\pi+2m\pi i t) = e^{2\pi i [n
(\alpha-1/2) + m(\beta-1/2)]}\psi^\mu(w)$}, the Green functions
for $\Psi$ can be written%
\begin{equation} \tag{\ref{psicorr}$'{'}$}
\vev{\Psi^\mu(w_1) \Psi^{\nu}(w_2)}
= \frac{\apr}{4\pi} G^{\mu\nu}
  \frac{\vartheta\left[\begin{smallmatrix} \alpha \\ \beta
                       \end{smallmatrix} \right]
               \bigl(\frac{w_1-w_2}{2\pi}|it\bigr)
         \vartheta'\left[\begin{smallmatrix} 1/2 \\ 1/2
                       \end{smallmatrix}\right](0|it)}%
        {\vartheta\left[\begin{smallmatrix} 1/2 \\ 1/2 
                       \end{smallmatrix} \right]
               \bigl(\frac{w_1-w_2}{2\pi}|it\bigr)
         \vartheta\left[\begin{smallmatrix} \alpha \\ \beta 
                        \end{smallmatrix}\right](0|it)},
\end{equation}
and the partition function is
\begin{equation} \tag{\ref{ppsi}$'$}
\vev{1} = \frac{\vartheta\left[\begin{smallmatrix}\alpha \\ \beta
                               \end{smallmatrix}\right]\bigl(0|it\bigr)^4}%
               {\eta(it)^4}.
\end{equation}

It is well known (see {\em e.g.\/}~\cite{jp})
that, on the annulus, the first nonvanishing
correlation function involves four pairs of fermions.  This gives 
the one-loop nonrenormalization theorem for
the $F^2$ and $F^3$ terms in the effective supersymmetric spacetime
action.
Because the effect of the $B$-field on the fermions can be
absorbed into the doubling trick, this continues to hold.%
\footnote{It is easy to check this, using equations~(\ref{ppsi}$'$)
and~(\ref{psicorr}$'{'}$), and the Jacobi fundamental formula for sums of
products of
$\vartheta$-functions~\cite[p.~467--468]{ww} of which the abstruse
identity is a special case.}
Thus, the
first nontrivial amplitude on the annulus involves four gauge bosons,
and the corresponding amplitude is
\begin{multline} \label{tocalc}
\sum_{\pm,\pm'} \frac{g_o^4}{4\apr^2} \, (2 \alpha')^4 \, T^a T^b T^c T^d
e^a_{1\mu} e^b_{2\nu} e^c_{3\rho} e^d_{4\sigma}
\vev{e^{i k_1 \cdot X}(w_1) e^{i k_2 \cdot X}(w_2) 
     e^{i k_3 \cdot X}(w_3) e^{i k_4 \cdot X}(w_4)}
\\* \times
\vev{\no{k_1{\cdot} \Psi \Psi^{\mu}}(w_1)
     \; \no{k_2{\cdot} \Psi \Psi^{\nu}}(w_2) 
     \; \no{k_3{\cdot} \Psi \Psi^{\rho}}(w_3)
     \; \no{k_4{\cdot} \Psi \Psi^{\sigma}}(w_4)}_{\pm \pm'}
\end{multline}
where we have factored the amplitude into fermionic and bosonic
factors, and suppressed the ghosts.  The sum is over spin structures.
The fermionic contribution is
easily found to be
\begin{multline} \label{fermcalc}
\sum_{\pm,\pm'}
\vev{\no{k_1{\cdot} \Psi \Psi^{\mu}}(w_1)
     \; \no{k_2{\cdot} \Psi \Psi^{\nu}}(w_2) 
     \; \no{k_3{\cdot} \Psi \Psi^{\rho}}(w_3)
     \; \no{k_4{\cdot} \Psi \Psi^{\sigma}}(w_4)}_{\pm \pm'}
\\
= 
\eta(it)^8 
k_{1\tau} k_{2\lambda} k_{3\alpha} k_{4\beta}\,
t^{\tau\mu\lambda\nu\alpha\rho\beta\sigma},
\end{multline}
where
$t^{\alpha_1\beta_1\alpha_2\beta_2\alpha_3\beta_3\alpha_4\beta_4}$ is
antisymmetric in \hbox{$\alpha_i\leftrightarrow\beta_i$} (for each $i$) and
is symmetric in \hbox{$(\alpha_i,\beta_i)\leftrightarrow(\alpha_j,\beta_j)$}.
It is convenient to define
\begin{subequations} \label{fermcalct}
\begin{align}
{\mathcal K} & \equiv  k_{1\alpha_1}e^a_{1\beta_1} k_{2\alpha_2}e^b_{2\beta_2}
k_{3\alpha_3}e^c_{3\beta_3} k_{4\alpha_4}e^d_{4\beta_4}
t^{\alpha_1\beta_1\alpha_2\beta_2\alpha_3\beta_3\alpha_4\beta_4}  \\
\intertext{which gives}
{\mathcal K} & = 
M_{1 \mu}{^\lambda} M_{2 \lambda}{^\alpha} M_{3 \alpha}{^\nu}
M_{4 \nu}{^\mu} - \frac{1}{4} M_{1 \mu \nu } 
M_{2}^{\nu \mu} M_{3 \tau \sigma}
M_{4}^{ \sigma \tau} + \text{2 permutations}.
\end{align}
\end{subequations}
In~\eqref{fermcalct}, \hbox{$M_{i \mu \nu} = e_{i \mu} k_{i \nu} - k_{i \mu}
e_{i \nu}$},
the indices are raised and lowered using the open string metric $G$, and 
the two permutations 
are given by changing the ordering from (1234) to $(1342)$ and 
$(1423)$ (see {\em e.g.}~\cite{jp}).

The bosonic contribution is identical to the tachyon correlation
function computed for the bosonic string in~\cite{dorn,lee,bcr,shen,lm,cn}.
Combining equations~\eqref{fermcalc} 
and~\cite[eq.~(2.15)]{lm} gives the total amplitude
\begin{multline} \label{pretotamp}
{\mathcal A} = -i \sqrt{\det G} \frac{g_o^4}{4 \apr^2} (2 \apr)^4 \,
(2\pi)^{p+1} \delta^{(p+1)} \Bigl(\text{\small $\sum_{q=1}^4$} k_q \Bigr) \,
{\mathcal K}
\\ \times
\int_0^\infty \frac{dt}{2t} (8 \pi^2 \apr t)^{-\frac{p+1}{2}} \,
\exp\left[\frac{k_\mu (\Theta G \Theta)^{\mu\nu} k_\nu}{8\pi\apr t}
- \frac{r^2 t}{2\pi \apr}\right]
\left[\prod_{q=1}^{4} \int_{0}^{2 \pi t} d \tau_q \right] \,
\Psi_1 \Psi_2 \Psi_{12}
\\ \times
\exp \left[-\frac{i}{2}\sum_{i<j} (k_i \times k_j) [\frac{\tau_{ij}}{\pi t}
- \ep(\tau_{ij})]
+\frac{i}{2} \sum_{y<z} (k_y \times k_z) [\frac{\tau_{yz}}{\pi t}
- \ep(\tau_{yz})] \right]
\end{multline}
where $\tau_{ij} = \tau_i-\tau_j$, $k = \sum_i k_i = -\sum_y k_y$ is
the nonplanar momentum,
$\epsilon(\tau)$ is the sign of $\tau$, and
\begin{subequations}
\begin{gather}\label{defPsis}
\begin{align} 
\Psi_1 &= \prod_{i<j} \abs{\psi_{ij}}^{2 \apr k_{i\mu} G^{\mu\nu} k_{j\nu}}, &
\Psi_2 &= \prod_{y<z} \abs{\psi_{yz}}^{2 \apr k_{y\mu} G^{\mu\nu} k_{z\nu}}, &
\Psi_{12} &= \prod_{i,y} (\psi^T_{iy})^{2 \apr k_{y\mu} G^{\mu\nu}k_{i\nu}},
\end{align} \\\label{defpsis}
\begin{align} 
\psi_{ij}  &= 2 \pi i \exp \left( -\frac{\tau_{ij}^2}{4 \pi t}\right)
\frac{\vt_{1}(i\frac{\tau_{ij}}{2 \pi}|it)}{\vt_{1}'(0|it)}, &
\psi_{iy}^T &= 2 \pi \exp \left( -\frac{\tau_{iy}^2}{4 \pi t}\right) \,
\frac{\vt_{2}(i\frac{\tau_{iy}}{2 \pi}| it)}{\vt_{1}'(0|it)}.
\end{align}
\end{gather}
\end{subequations}
Here, $i$ and $y$ are indices which run over the vertex operators on
the two boundaries, and $q$ runs over all vertex operators.  
The quantity ${\mathcal K}$ was given in~\eqref{fermcalct},
and we have suppressed Chan-Paton factors. 
We also have introduced a transverse distance $r$
between the parallel branes. It gives a mass
\begin{equation}
m = \frac{r}{2 \pi \apr}
\end{equation}
to the ground states of open strings stretching between the branes and  
acts as an infrared cutoff in the world-volume Yang-Mills theory.

The exponential factor
\begin{equation} \label{importf}
\exp\left[\frac{1}{8 \pi \alpha' t} 
k_\mu (\Theta G \Theta)^{\mu\nu} k_\nu \right]
= \exp\left[- \frac{(\Delta x)^2}{8 \pi \alpha' t} 
   \right], \qquad \Delta x^{\mu} \equiv \Theta^{\mu \nu} k_{\nu},
\end{equation}
in~\eqref{pretotamp}, which was interpreted in~\cite{lm} as the
manifestation of a
stretched string in the bosonic string theory, and was responsible for
IR/UV mixing
in those theories, survives the field theory limit in the
superstring as well.  Thus, the observations made in~\cite{lm} apply also
here.

\section{One-loop Amplitude in Noncommutative 
${\mathcal N} =4$ SYM} \label{sec:ft}

The one-loop amplitude for noncommutative 
${\mathcal N} = 4$  super-Yang-Mills theory can be obtained
from~\eqref{pretotamp} by taking $\apr\rightarrow0$ with $\apr t$ fixed.
While our main focus is on the 3+1-dimensional ${\mathcal N} =4$
noncommutative SYM theory, our results
apply to generic $p$, and we will keep our discussion general.
To take the field theory limit, we have been implicitly considering
only magnetic $B$; however, the results of section~\ref{sec:calc} are
completely general, holding also for electric $B$, and so can be used
in the context of the NCOS~\cite{gmms,sst,gmss,bb,km}.
Note that as $t \rightarrow \infty$,
\begin{align}
\psi_{ij} &= -2 \sinh \bigl(\frac{\tau_{ij}}{2}\bigr)
\exp \bigl(-\frac{\tau_{ij}^2}{4 \pi t}\bigr),&
\psi_{iy}^T &= 2 \cosh \bigl(\frac{\tau_{iy}}{2}\bigr)
\exp \bigl(-\frac{\tau_{iy}^2}{4 \pi t}\bigr).
\end{align}
To isolate the $t$ dependence we rescale $\tau$ so that
$0<\tau<1$. Let $T = 2 \pi m^2 \apr t$ and recall that 
$g_{\text{YM}} = g_o/\sqrt{2 \apr}$. 
This gives the field theory amplitude
\begin{multline} \label{totamp}
{\mathcal A} = -i 2^4 \pi^{\frac{p+1}{2}} \sqrt{\det G}
 g_{\text{YM}}^4
\,\delta^{(p+1)} \Bigl(\text{\small $\sum_{q=1}^4$} k_q \Bigr)\,
{\mathcal K} \, m^{p-7}
\int_0^\infty \frac{dT}{2T} \, T^{\frac{7-p}{2}}
\exp\left[-\frac{m^2 (\Delta x)^2}{4 T} 
-  T \right]
\\* \times
\left[\prod_{q=1}^{4} \int_{0}^{1} d \tau_q \right]
\exp \left[ -\frac{i}{2}\sum_{i<j} (k_i \times k_j) [2\tau_{ij}
- \ep(\tau_{ij})]
+\frac{i}{2} \sum_{y<z} (k_y \times k_z) [2\tau_{yz}
- \ep(\tau_{yz})] \right]
\\* \times
\exp \left [ 
  T \sum_{p<q} \frac{k_{p\mu} G^{\mu\nu} k_{q\nu}}{m^2}
  \bigl[ \abs{\tau_{pq}} - \tau_{pq}^2 \bigr] \right].
\end{multline}

Later we shall examine the amplitude~\eqref{totamp}
in the kinematic region \hbox{$k_{p\mu} G^{\mu\nu} k_{q\nu} \ll m^2$}, with
$k_p \Theta k_q$ remaining finite [equation~\eqref{kin}].%
\footnote{We can also define $T = \frac{2 \pi \apr t}{(\Delta x)^2}$ and  
consider an expansion in terms of $k^2 (\Delta x)^2$. Either expansion gives 
the same leading order result, with which we are mostly concerned.}
In this case the 
last factor of~\eqref{totamp} has the expansion
\begin{equation} \label{ske}
\left[ 1 + T \sum_{p<q} \frac{k_{p\mu} G^{\mu\nu} k_{q\nu}}{m^2}
\bigl[ \abs{\tau_{pq}} - \tau_{pq}^2 \bigr] + \ldots \right].
\end{equation}

Let us first look at the planar contribution. In this case $k=0$ and 
$\Delta x =0$. Also, 
$\sum_{i<j} (k_i \times k_j) \tau_{ij} = 0$ by
momentum conservation; thus the phase factor in~\eqref{totamp} is just
\begin{equation} \label{planarphase}
\exp \left[\frac{i}{2}\sum_{p<q} (k_p \times k_q) \ep(\tau_{pq}) \right].
\end{equation}
This turns the amplitude 
\begin{equation} 
{\mathcal A} \sim \frac{1}{m^4} \int d^4 x \Tr F^4 
+ \frac{1}{m^2} \int d^4 x \Tr F F  DF DF + \cdots
\end{equation} 
of the commutative theory into the
\begin{equation} 
{\mathcal A} \sim \frac{1}{m^4} \int d^4 x \Tr(F\ast F\ast F\ast F)  
+ \frac{1}{m^2} \int d^4 x  
\Tr(F \ast F \ast DF \ast DF ) + \cdots
\end{equation} 
of the noncommutative theory.
(Here we have been schematic and have not included the Lorentz indices.)

The story for nonplanar diagrams is more complicated.
There are essentially two new features for nonplanar diagrams 
in~\eqref{totamp} compared to $B=0$ and the planar diagram. The first is 
the appearance, in the integral over proper
time $T$, of $(\Delta x)^2$, in a manner 
which suppresses the contribution to the amplitude from small $T$
(short-distance).
The second new feature is that in the integration of the vertex
operator coordinates, there are additional phase factors
\begin{equation} 
\exp \left[- i \sum_{i<j} (k_i \times k_j) (\tau_{i} - \tau_{j}) 
\right].
\end{equation}
Both features already appeared  in the bosonic cases~\cite{dorn,bcr,lm}. 
Here with a large number of supersymmetries present, their effects 
can be isolated and  subjected to a thorough analysis.

\subsection{Stretched strings and closed strings}

The factor $\exp \left[- \frac{ m^2 (\Delta x)^2}
{4 T} \right]$, in equation~\eqref{totamp}, was interpreted
in~\cite{lm} as the effect
of a stretched string. More explicitly, when $B \neq 0$, 
an open string has 
a nonzero length, as described in the introduction, the effect of which 
is to cut open an otherwise closed first-quantized particle loop.
The exponential factor may be understood as the amplitude for diffusion
of a mass $m$ particle over a distance $\abs{\Delta x}$. When, for
\hbox{$B=0$}, the loop
integral is divergent at short distances (small $T$), 
the stretched string effect regularizes the divergence. However, taking
$\Delta x$ to zero will recover the original UV divergence. This is the origin
of the IR/UV mixing discussed in~\cite{mrs}. In theories for which the
commutative
loop integral is convergent in the UV region, we would expect that the
$\Delta x\rightarrow 0$ limit can be achieved in a continuous way,
and that IR singularities are absent.
It might be expected that the effect of the stretched string is mild 
in these cases. However, as we shall see below, it suppresses the
amplitude exponentially at large external $k$.

In the low energy expansion in terms of $\frac{k_{p} \cdot k_{q}}{m^2}$
the integration over $T$ can be performed exactly for each term in the series 
and gives~\cite[eq.~8.432.6]{gr}
\begin{equation} \label{Tint}
I_n = m^{p-7} \int_0^\infty \frac{dT}{2T} T^{\frac{7-p}{2} + n}
e^{-\frac{m^2 (\Delta x)^2}{4 T}  - T}
= \left(\frac{m \abs{\Delta x}}{2}\right)^n 
\left(\frac{2 m }{\abs{\Delta x}}\right)^{\frac{p-7}{2}}
K_{\frac{7-p}{2} + n} (m \abs{\Delta x}),
\end{equation}
where $K$ is the modified Bessel function. Recall that 
$K_{-\nu} (z) = K_\nu (z)$ and as $z \rightarrow 0$,
\begin{equation}
K_{\nu} (z) = \begin{cases} 
  \frac{\Gamma(\nu)}{2}\bigl(\frac{2}{z}\bigr)^{\nu},& \nu>0, \\
            - \ln \frac{z}{2},& \nu=0. \end{cases}
\end{equation}
For generic $p$ and $n$, as $\Delta x \rightarrow 0$, we have
\begin{equation} \label{smxb}
I_{n} \sim 
\begin{cases}
\frac{\Gamma\bigl(n+\frac{7-p}{2}\bigr)}{2} m^{p-7},
   & n + \frac{7-p}{2} > 0, \\
-\log m \abs{\Delta x}, & p=7, n=0, \\
\frac{\Gamma\bigl(\frac{7-p}{2}\bigr)}{2}(\frac{2}{\abs{\Delta x}})^{p-7}, 
   & p=8, 9; n=0, \\
-m^2 \log (m \abs{\Delta x}), & p=9, n=1.
\end{cases}
\end{equation}  
When $n+\frac{7-p}{2}>0$, the integral in~\eqref{Tint} is convergent  
as $\Delta x =0$;
thus, as $\Delta x \rightarrow 0$, $I_n$ reduces to the 
standard results for $B=0$. In  the other cases, logarithmic,
linear and quadratic divergences, respectively for 8, 9 and 10-dimensional
SYM theory, are reflected in the singular dependence on $\Delta x$ in the
amplitude. 

For $m \abs{\Delta x}  \gg 1$---{\em i.e.\/}\ the
large external momentum limit---
\begin{equation} \label{diff}
I_n \sim \exp [- m \abs{\Delta x}].
\end{equation}
The amplitude is exponentially suppressed at large $k$. The behaviour
in~\eqref{diff} is easily understood from the stretched string picture 
in which a particle of mass $m$ diffuses a distance $\abs{\Delta x}$.
This large momentum suppression is 
reminiscent of the asymptotic behaviour of correlation functions 
observed in the supergravity side~\cite{mr} and the high temperature 
suppression of the partition function for nonplanar diagrams 
of ${\mathcal N} =4$ noncommutative SYM~\cite{texas}, although the
details are somewhat different.

We note that for $n=0$---{\em i.e.\/}\ the leading term in the 
expansion~\eqref{ske}---there is also 
an interpretation for~\eqref{Tint} in terms of the lowest 
closed string modes. Defining $M^2 = (\frac{\Delta x}{2 \pi \apr})^2 $
and $\tilde{d} = 9-p$, the codimension of the brane, we can rewrite 
equation~\eqref{Tint} as (notice that $M r = m \abs{\Delta x}$ ) 
\begin{equation} \label{propeq}
I_0 = (2 \pi \apr)^{7-p} 2\pi^{\frac{9-p}{2}}
\int \frac{d^{\tilde{d}} q}{(2 \pi)^{\tilde{d}}}
\frac{e^{i q \cdot r}}{q^2 + M^2}.
\end{equation}
The right hand side of equation~\eqref{propeq} has the interpretation 
of a massive particle of mass $M$  propagating between
the branes.  In contrast with the $B=0$ case 
($\Delta x \rightarrow 0$ in~\eqref{Tint}), for which the propagator 
is
\hbox{$m^{p-7}=\bigl(\frac{2\pi\apr}{r}\bigr)^{7-p}$}
corresponding to a
virtual mediating massless closed string, now the propagating particle
has a mass $M = \frac{\abs{\Delta x}}{2 \pi \apr}$. 
This result can be understood by noting that
the intermediate closed string has momentum $k_\mu$ (defined below 
equation~\eqref{pretotamp}) along the brane 
direction from momentum conservation and
therefore an effective mass 
\begin{equation} \label{mcl}
M^2 = -k_\mu g^{\mu\nu} k_\nu = -k_\mu G^{\mu\nu} k_\nu +
\left(\frac{1}{2\pi\apr}\right)^2 k_\mu (\Theta G \Theta)^{\mu\nu} k_\nu 
\sim \left(\frac{\Delta x}{2\pi\apr}\right)^2,
\end{equation}
where in the last step we have recalled equation~\eqref{kin}.
Physically, this means that due to the different scaling between the
closed and
open string metrics, the low energy limit for the brane world-volume theory
does not correspond to the low energy limit for the bulk closed string
modes. 

\skipthis{
The mass~\eqref{mcl} for the closed string modes can also be understood
heuristically as follows\footnote{The following picture was developed 
with Miao Li.}. 
Nonplanar processes can be imagined as  incoming open strings 
joining their ends together to form an  intermediate 
closed string at one boundary and then splitting into outgoing open strings 
at the other boundary.  Since in the presence of 
$B$ field, each incoming
open string has a length $\Theta k_i$, altogether all
the incoming
open strings have a total displacement of
\begin{equation}
\Delta x^{\mu} = \sum_{i} \Theta^{\mu \nu} k_{i \nu} 
= \Theta^{\mu\nu} k_{\nu}.
\end{equation}
Thus the intermediate closed string has a macroscopic length of 
$|\Delta x|$, giving rise to a mass $M = \frac{|\Delta x|}{2 \pi \apr}$. 
}%
It is interesting to compare the IR/UV relation between the open and closed
string pictures. For $p \leq 6$, SYM theory is IR divergent when $m=0$; this
is reflected in the UV divergence, for codimension $\tilde{d} \geq 3$, of
the transverse propagator on the closed string side, as we reduce the
separation of the
branes to zero. For $p=8,9$, the SYM theory is linearly and quadratically 
UV divergent while the transverse propagator is IR divergent ($M \rightarrow
0$) in codimension $\tilde{d}=1,0$. At $p=7$ SYM theory is both UV and IR
divergent and the closed string propagator in codimension $2$ 
is both IR and UV divergent.

The UV/IR relations in equation~\eqref{smxb} for $p=7,8,9$ are
precise realizations
of closed string idea proposed in~\cite{mrs,rs} to explain IR/UV mixing 
in noncommutative theories. 
However we note that that picture seems to  apply only for the $n=0$
term, which is protected by a nonrenormalization theorem from contributions
of massive string modes~\cite{dkps}. 
For higher order terms, in principle, there can  be an 
infinite number of massive closed string modes involved. For example, 
the $p=9, n=1$ term is logarithmically divergent as $\Delta x \rightarrow 0$, 
but there 
is no propagation of a codimension two closed string to explain it.%
\footnote{The closed string proposal also suffers problems at two
loops.\cite{klp}}

\subsection{Beyond the $\ast$-product} \label{beyond}
Now we would like to focus on the leading order ($ n=0$) terms, and attempt to
write down an off-shell gauge invariant one-loop effective
action. 

Due to the presence of the additional phase factors,
\begin{equation} \label{addph}
\exp \left[- i \sum_{i<j} (k_i \times k_j) (\tau_{i} - \tau_{j}) 
\right],
\end{equation}
in the integrals over $\tau_q$ in~\eqref{totamp}, the product 
between the insertions on the same boundary is neither an
ordinary nor a $\ast$-product, as it was for the $B=0$, and planar cases,
respectively.
In particular and interestingly, the nonplanar terms in the action are not
$\ast$-product generalizations of the familiar ones.

For the nonplanar diagram with two vertex operators on each boundary
(we label $1,2$ on one boundary and $3,4$ on the
other), we
find that the $\tau$-integrations give rise (including the Chan-Paton factors) 
to a factor of
\begin{equation} \label{tot2nonplanar}
\Tr(T^a T^b) \Tr (T^c T^d) 
\frac{\sin\bigl(\frac{k_1 \times k_2}{2}\bigr)}{\frac{k_1\times k_2}{2}}
\frac{\sin\bigl(\frac{k_3 \times k_4}{2}\bigr)}{\frac{k_3\times k_4}{2}}.
\end{equation}
Note that~\eqref{tot2nonplanar} is
finite as $k_1 \times k_2 \rightarrow 0$ and/or $k_3 \times k_4
\rightarrow 0$.

The other type of nonplanar diagram has $1,2,3$ on one boundary and
$4$ on the other.  Then, for the ordering $\tau_1>\tau_2>\tau_3$,
\begin{multline} \label{nonplanar31}
\int_0^1 d\tau_1 \int_0^{\tau_1} d\tau_2 \int_0^{\tau_2} d\tau_3 \;
e^{\frac{i}{2} (k_1\times k_2 + k_1\times k_3 + k_2 \times k_3)}
e^{-i (k_1\times k_2) \tau_{12} - i(k_1\times k_3) \tau_{13}
   -i (k_2\times k_3) \tau_{23}}
\\
= \frac{e^{\frac{i}{2} (k_1\times k_2 + k_3 \times k_4)}}%
       {(k_1\times k_4)(k_3\times k_4)}
+ 2 \frac{e^{\frac{i}{2} k_2\times k_3} \sin \bigl(\frac{k_1\times k_4}{2}
            \bigr)}{(k_1\times k_4)^2 (k_2\times k_4)}
+ 2 \frac{e^{\frac{i}{2} k_1\times k_2} \sin \bigl(\frac{k_3\times k_4}{2}
            \bigr)}{(k_2 \times k_4)(k_3 \times k_4)^2}.
\end{multline}
This is nonsingular even when $k_i \times k_j \rightarrow 0$.
Including all the orderings, as well as the minus sign that comes
with the odd-number of Chan-Paton matrices on the $\sigma=\pi$
boundary, gives%
\skipthis{\footnote{The sign comes from converting the second trace from the
antifundamental to the fundamental representation.}}%
\begin{smaleq}
\begin{multline} \label{totnonplanar31}
-\Tr(T^b T^a T^c) \Tr(T^d) \left\{
\frac{e^{\frac{i}{2}(k_1\times k_2 + k_1\times k_3 +k_2\times k_3)}}%
     {[k_3 \times (k_1+k_2)][k_1\times (k_2+k_3)]}
+ \frac{e^{\frac{i}{2}(k_2\times k_3 - k_1\times k_2 -k_1\times k_3)}}%
     {[k_1 \times (k_2+k_3)][k_2\times (k_1+k_3)]}
\right. \\* \left.
+ \frac{e^{\frac{i}{2}(k_1\times k_2 - k_1\times k_3 -k_2\times k_3)}}%
     {[k_2 \times (k_1+k_3)][k_3\times (k_1+k_2)]}
\right\} \\*
- \Tr(T^a T^b T^c) \Tr(T^d) \left\{
\frac{e^{\frac{i}{2}(k_2\times k_3 + k_1\times k_3 -k_1\times k_2)}}%
     {[k_3 \times (k_1+k_2)][k_2\times (k_1+k_3)]}
+ \frac{e^{\frac{i}{2}(k_1\times k_3 + k_1\times k_2 -k_2\times k_3)}}%
     {[k_2 \times (k_1+k_3)][k_1\times (k_2+k_3)]}
\right. \\* \left.
+ \frac{e^{-\frac{i}{2}(k_1\times k_2 + k_1\times k_3 +k_2\times k_3)}}%
     {[k_1 \times (k_2+k_3)][k_3\times (k_1+k_2)]}
\right\}.
\end{multline}
\end{smaleq}%
\skipthis{The reader might be concerned that, for the $1\rightarrow3$
diagram, the two curly brackets should be exchanged
in~\label{totnonplanar31}, because $\Tr_{\Bar{N}} T^a T^b T^c = -\Tr_N
T^c T^b T^a$.  However, this is undone by the opposite sign of the
phase factors on the $\sigma=\pi$ boundary.}%
When the gauge group is abelian, equation~\eqref{totnonplanar31} can
be simplified to%
\footnote{This is not manifestly symmetric in the three vertex
operators on one boundary, but follows from the manifestly
symmetric expression~\eqref{totnonplanar31}.}
\begin{equation} \label{u1totnonplanar31}
-\frac{\sin\bigl(\frac{k_2\times k_3}{2}\bigr) 
\sin\bigl(\frac{k_1\times (k_2+ k_3)}{2}\bigr)}%
{\frac{(k_1+k_2)\times k_3}{2} \frac{k_1\times(k_2+k_3)}{2}}
- \frac{\sin\bigl(\frac{k_1\times k_3}{2}\bigr) 
\sin\bigl(\frac{k_2\times (k_1+ k_3)}{2}\bigr)}%
{\frac{(k_1+k_2)\times k_3}{2} \frac{k_2\times(k_1+k_3)}{2}}.
\end{equation}

If we have $N$ vertex operators on the same boundary, the general product
structure between the operators is given by
\begin{equation} \label{genpro}
\Tr\left[T^{a_1}\dots T^{a_N}\right]
\left[\prod_{q=1}^{N} \int_{0}^{\tau_{q-1}} d \tau_q \right]
\exp \left[ -\frac{i}{2}\sum_{i<j} (k_i \times k_j) [2\tau_{ij}
- \ep(\tau_{ij})] \right]; \quad \tau_0 \equiv 1,
\end{equation}
plus permutations.
\skipthis{The reader who is worried about the opposite sign of the
phase factors on the $\sigma=\pi$ boundary, should note that the trace
is taken in the antifundamental on that boundary, which reverses the
ordering.  This is the same effect as changing the sign of $\tau_{ij}$;
that is, the opposite phase factor goes with the opposite ordering
(and vice versa) so indeed we have the same
structure,~\eqref{genpro}, on the $\sigma=\pi$ boundary as well.}%

To lowest order in $k^2/m^2$, the four gluon scattering amplitudes 
are given by
\begin{multline} \label{two}
{\mathcal A}_{2 \rightarrow 2} 
 = -i 2^4 \pi^{\frac{p+1}{2}} \sqrt{\det G}
 g_{\text{YM}}^4
\,\delta^{(p+1)} \Bigl(\text{\small $\sum_{q=1}^4$} k_q \Bigr)\,
\Tr(T^a T^b) \Tr (T^c T^d)\; {\mathcal K} \, \\* \times
\left(\frac{\Delta x}{2m}\right)^{\frac{7-p}{2}} 
K_{\frac{7-p}{2}}(m \Delta x) \, 
\frac{\sin\bigl(\frac{k_1 \times k_2}{2}\bigr)}{\frac{k_1\times k_2}{2}}
\frac{\sin\bigl(\frac{k_3 \times k_4}{2}\bigr)}{\frac{k_3\times k_4}{2}},
\end{multline}
where ${\mathcal K}$ is given by~\eqref{fermcalct}.
Similarly (for simplicity, we only give the $U(1)$ result in this case),
\begin{multline} \label{oneth}
{\mathcal A}_{1 \rightarrow 3} 
 = i 2^4 \pi^{\frac{p+1}{2}} \sqrt{\det G}
 g_{\text{YM}}^4
\,\delta^{(p+1)} \Bigl(\text{\small $\sum_{q=1}^4$} k_q \Bigr)\,
{\mathcal K} \, \\* \times
\left(\frac{\Delta x}{2m}\right)^{\frac{7-p}{2}} 
K_{\frac{7-p}{2}}(m \Delta x) \, 
\left[ 
\frac{\sin\bigl(\frac{k_2\times k_3}{2}\bigr) 
\sin\bigl(\frac{k_1\times (k_2+ k_3)}{2}\bigr)}%
{\frac{(k_1+k_2)\times k_3}{2} \frac{k_1\times(k_2+k_3)}{2}}
+ \frac{\sin\bigl(\frac{k_1\times k_3}{2}\bigr) 
\sin\bigl(\frac{k_2\times (k_1+ k_3)}{2}\bigr)}%
{\frac{(k_1+k_2)\times k_3}{2} \frac{k_2\times(k_1+k_3)}{2}}
\right].
\end{multline}
It is easy to see that the above amplitudes are gauge invariant on-shell;
{\em i.e.\/}\ ${\mathcal A} = 0$, when we take $e_i \propto k_i$.

Now we would like to extend the above on-shell amplitudes to the
off-shell effective
action. The standard way of doing this is to take 
$M_{\mu \nu} = e_\mu k_\nu - e_\nu k_\mu \rightarrow 
- \frac{ i}{g_{YM}} F_{\mu \nu}$
with $F_{\mu \nu}$ given by the tree-level expression
\begin{equation} \label{ncfs}
F_{\mu \nu} = \p_{\mu} A_{\nu} - \p_{\nu} A_{\mu} 
- i A_\mu \ast A_\nu
+ i A_{\nu} \ast A_{\mu}
\end{equation}

Combining these expressions, we find that the 1-loop effective action
contains the terms%
\footnote{Using equation~\eqref{fermcalct}, we can write {\em
e.g.\/} (up to a total derivative if $\ast$-products are used)
\begin{multline*}
t^{\mu\nu\rho\sigma\lambda\tau\alpha\beta} \Tr \left( 
F_{\mu\nu}(x) F_{\rho\sigma}(x) F_{\lambda\tau}(x)   
F_{\alpha\beta}(x)] \right) \\*
= 2^4 \Tr \left( 2 F_{\mu \rho} F_{\nu}{^\rho} F^{\mu \sigma} F^{\nu}{_\sigma}
+ F_{\mu}{^\rho} F_{\rho}{^\nu} F_{\nu}{^\sigma} F_{\sigma}{^\mu}
-\frac{1}{2} F_{\mu \rho} F^{\mu \rho} F_{\nu \sigma} F^{\nu \sigma}
-\frac{1}{4} F_{\mu \rho} F_{\nu \sigma} F^{\mu \rho} F^{\nu \sigma}
\right)
\end{multline*}
}

\begin{subequations} \label{ncf4}
\begin{equation} \label{usencf4}
\begin{split}
\Gamma_{\text{1-loop}} & = -\frac{1}{4! (4\pi)^{\frac{p+1}{2}}}
\int d^{p+1} x \, \sqrt{\det G} \,
t^{\mu\nu\rho\sigma\lambda\tau\alpha\beta} 
\\* & \mspace{-40mu} \times 
\biggl\{
\frac{N_2}{2} \Gamma\bigl(\frac{7-p}{2}\bigr) \frac{1}{m^{7-p}}
\Tr_{U(N_1)}[F_{\mu\nu}(x)\ast F_{\rho\sigma}(x)\ast F_{\lambda\tau}(x) \ast 
    F_{\alpha\beta}(x)]
\\* & \mspace{-40mu}
+ \left.
3\Tr_{U(N_1)}[F_{\mu\nu}(x) \ast' F_{\rho\sigma}(x)]\, 
I_{\frac{7-p}{2}}\Bigl(m,\sqrt{\lvec[\mu]{\p}\bigl(\Theta G 
      \Theta\bigr)^{\mu\nu}\rvec[\nu]{\p}}\Bigr)
\Tr_{U(N_2)}[F_{\lambda\tau}(x) \ast' F_{\alpha\beta}(x)] 
\right. \\*
& \mspace{-40mu} - \biggl. 
4 \Tr_{U(N_1)}
   \bigl(\ast_3[F_{\mu\nu}(x),F_{\rho\sigma}(x),F_{\lambda\tau}(x)]\bigr)
I_{\frac{7-p}{2}}\Bigl(m,\sqrt{-\p_\mu \bigl(\Theta G \Theta\bigr)^{\mu\nu}
     \p_\nu} \Bigr)
\Tr_{U(N_2)}(F_{\alpha\beta}(x))
\biggr\} 
\\* & \mspace{-40mu} + (N_1 \rightarrow N_2)
\end{split}
\end{equation}
where $N=N_1+N_2$ is the total number of D-branes with $N_1$ and
$N_2$ D-branes on each boundary; the factor of $N_{1,2}$ appears in the planar
diagram from tracing
over the Chan-Paton factors on the empty boundary; the traces are
taken in the fundamental of the indicated subgroup;
\begin{equation} \label{defImp}
I_{n}(m,x) \equiv \left(\frac{x}{2m}\right)^n K_n(m x);
\end{equation}
\end{subequations}
and we have defined, from~\eqref{tot2nonplanar},
the $\ast'$-product~\eqref{def*'},
and from~\eqref{totnonplanar31}, the $\ast_3$-%
ternary operation.  For an abelian gauge group, the $\ast'_3$-ternary
operation
\begin{multline} \label{ternary}
\ast'_3[f(x),g(x),h(x)] \\* \equiv 
\left[
\frac{\sin\bigl(\frac{\partial_2\times \partial_3}{2}\bigr) 
\sin\bigl(\frac{\partial_1\times (\partial_2+ \partial_3)}{2}\bigr)}%
{\frac{(\partial_1+\partial_2)\times \partial_3}{2} 
\frac{\partial_1\times(\partial_2+\partial_3)}{2}}
+ \frac{\sin\bigl(\frac{\partial_1\times \partial_3}{2}\bigr) 
\sin\bigl(\frac{\partial_2\times (\partial_1+ \partial_3)}{2}\bigr)}%
{\frac{(\partial_1+\partial_2)\times \partial_3}{2}
\frac{\partial_2\times(\partial_1+\partial_3)}{2}}\right]
\evalat{f(x_1) g(x_2) h(x_3)}{x_i=x},
\end{multline}
defined following~\eqref{u1totnonplanar31}, gives the same result as
the similarly defined $\ast_3$-ternary operation.
For the general product~\eqref{genpro} we can define a $\ast_n$
$n$-ary operation which we will not write down explicitly.

We should note that (for $p<7$) equations~\eqref{Tint} and~\eqref{smxb}
imply that the power expansion of \hbox{$I_{\frac{7-p}{2}}(m,x)$} contains
only nonnegative, even integer powers of $x$; thus
equation~\eqref{usencf4} contains only nonnegative integer powers of
derivatives, and
so is well-defined.  The limit $\Theta\rightarrow0$ is also
well-defined---for example, all the $\ast_n$ $n$-ary operations become
ordinary products---and recovers the nonabelian analogue
of~\eqref{kthree} given in {\em e.g.\/}~\cite{chtseyt} and references therein.

We have some remarks regarding equation~\eqref{ncf4}:
\begin{itemize}
\item The $\ast'$-product was also
found by Garousi in~\cite{mg}. When considering the scattering
amplitudes of one closed string with two open string modes,%
\footnote{These disk diagrams were also considered
in~\cite{hkll}.} 
he found
that the 
$\ast'$-product should be used to multiply the massless open string fields.
\item The nonplanar part of equation~\eqref{ncf4} is not 
gauge-invariant, off-shell. To be specific, we focus on the second 
line of~\eqref{usencf4}; the
symmetry of the expression---and particularly that of the $\ast'$-product---%
implies that this term transforms into
\begin{multline} \label{gtncf4}
6 i \Tr_{U(N_1)}[(\Lambda \ast F_{\mu\nu} - F_{\mu\nu} \ast \Lambda) \ast'
F_{\rho\sigma}]\,
I_{\frac{7-p}{2}}\Bigl(m,\sqrt{\lvec[\mu]{\p}\bigl(\Theta G 
      \Theta\bigr)^{\mu\nu}\rvec[\nu]{\p}}\Bigr)
\Tr_{U(N_2)}[F_{\lambda\tau} \ast' F_{\alpha\beta}]
\\ + (N_1\leftrightarrow N_2).
\end{multline}
At $B=\Theta=0$, the $\ast$- and $\ast'$-products reduce to the
ordinary product, and so the expression~\eqref{gtncf4} vanishes by
cyclicity of the trace and the symmetry properties of the tensor $t$.
For the planar contribution [the first term
of~\eqref{usencf4}] the associativity of the $\ast$-product, combined
with cyclicity of the trace, causes that
term to be gauge invariant.  Generically the total 1-loop
effective action is not gauge invariant.
\item In the commutative limit, $\Tr F^2$ and $\Tr F^3$  (with
appropriate
Lorentz contractions) are gauge invariant operators,
in~\eqref{usencf4}, which
couple to closed string modes in the bulk. They are also observables  
used in the AdS/CFT correspondence. However 
in the noncommutative theory, 
the corresponding operators in~\eqref{ncf4} are no longer gauge invariant. 
This is hardly surprising, since there are no gauge invariant local operators
in the theory~\cite{gross,dr,ghi}. In~\cite{dr,ghi},  
gauge invariant observables in NCSYM corresponding to supergravity 
modes~\cite{hi,mr} were constructed using open Wilson 
lines~\cite{iikk,ambjonf,amns2}.%
\footnote{Specifically, although Wilson loops are no longer gauge
invariant in a noncommutative field theory, the references show that
if a Wilson loop of definite momentum is ``cut open'' so that its
endpoints are separated by
precisely the displacement of the stretched
string, then the resultant (nonlocal) Wilson line is gauge
invariant.}
It is natural to wonder whether it is possible to replace~\eqref{ncf4}
by a gauge invariant version in terms of these observables or their 
generalizations. Our efforts in this direction have not yielded a positive 
answer.
\item $\ast_3'$ generalizes the $\ast'$-product above.
In fact, the products in equations~\eqref{totnonplanar31}
and~\eqref{u1totnonplanar31} do not appear to separate into pairs---%
this may be related to the fact that 
the $\ast'$-product is not associative.%
\footnote{The nonassociativity also means that we cannot gauge, in the sense
of~\cite{kont,fed} (see also~\cite{ncrev}), the $\ast'$-product to
the ordinary product.}
Thus, it seems more natural to talk about a generalized
$\ast_n$ $n$-ary {\em operation\/} rather than a $\ast$-product.  
For example, $\ast_3'$ defined in~\eqref{ternary} may be called 
a ternary operation.
\item Our above discussions have been restricted to the leading term 
in~\eqref{totamp}. For subleading terms the $\tau$-integration 
will generate a much more complicated product pattern. For example the
second term in~\eqref{ske} gives rise to
\begin{multline} 
\left[\prod_{q=1}^{4} \int_{0}^{1} d \tau_q \right]
\exp \left\{ -\frac{i}{2}\sum_{i<j} (k_i \times k_j) \bigl[2\tau_{ij}
- \ep(\tau_{ij})\bigr]
\right. \\* \left.
+\frac{i}{2} \sum_{y<z} (k_y \times k_z) \bigl[2\tau_{yz} 
- \ep(\tau_{yz})\bigr] \right\}
\sum_{p<q}   \bigl[ \abs{\tau_{pq}} - \tau_{pq}^2 \bigr] 
\end{multline}
Thus it is a generic phenomenon that, for nonplanar diagrams, the 
amplitudes are no longer expressible in terms of $\ast$-products alone,
and may not be extended off-shell in a gauge invariant way 
using~\eqref{ncfs}.
\end{itemize}

\section{Discussion and Conclusions} \label{sec:conc}

We have shown that for noncommutative ${\mathcal N}=4$ SYM at one-loop, 
just as for the commutative theory, there is no contribution to the $F^2$ 
and $F^3$ terms at one-loop (see also~\cite{mst}). This simply results from 
properties of fermion correlation functions on the worldsheet. We also
expect that the planar part of the $F^4$ term is not renormalized 
beyond one-loop as in the commutative case~\cite{dinesei}, but this is much 
less clear for the nonplanar part. 

We have written down an off-shell effective action which reproduces
the on-shell amplitudes~\eqref{two} and~\eqref{oneth}. 
This na\"{\i}ve off-shell extension,
equation~\eqref{ncf4},
is not gauge invariant.%
\footnote{One might hope to use the methods
of~\cite{iikk,ikk,ambjonf,amns2,dr,ghi} to write down a gauge
invariant off-shell extension, but we have not succeeded in doing so.}
Though equation~\eqref{ncf4} is only the first term in an expansion in
$k{\cdot}G{\cdot}k$---while exact in $\Theta$---from the last remark 
at the end of section~\ref{beyond} we expect that 
the lack of gauge invariance is generic for the higher order terms 
in the expansion~\eqref{ske}.

The reason may be attributed to the fact that even in the field theory limit,
the theory is stringy, as a result of the stretched string effect.
Recall that the worldsheet correlators for $X$ can be written as%
\footnote{Here we take a symmetric form of the propagators with respect
to the two boundaries, with $0\leq\tau_i<2 \pi t$; see appendix~\ref{sec:prop}.}
\begin{subequations} \label{wspro}
\begin{align} \label{wspro00}
{\mathcal G}^{\mu\nu}(i\tau_i,i\tau_j) & = - \alpha' G^{\mu \nu} \ln \abs{
\vartheta_{1}\bigl(\frac{\tau_{ij}}{2 \pi t}|\frac{i}{t}\bigr)}^2
- \frac{i}{2} \Theta^{\mu \nu} 
\left[\epsilon (\tau_{ij} ) - \frac{\tau_{ij}}{\pi t} \right]; \\ 
\label{wspro11}
{\mathcal G}^{\mu\nu}(\pi + i\tau_x,\pi + i\tau_y) & =
- \alpha' G^{\mu \nu} \ln \abs{
\vartheta_{1}\bigl(\frac{\tau_{xy}}{2 \pi t}|\frac{i}{t}\bigr)}^2
+ \frac{i}{2} \Theta^{\mu \nu} 
\left[\epsilon (\tau_{xy} ) - \frac{\tau_{xy}}{\pi t} \right]; \\ 
\label{wspro01}
{\mathcal G}^{\mu\nu} (i\tau_i, \pi + i\tau_x) & =
- \alpha' G^{\mu \nu} \ln \abs{
\vartheta_{4}\bigl(\frac{\tau_{ix}}{2 \pi t}|\frac{i}{t}\bigr)}^2
+ \frac{(\Theta G \Theta )^{\mu \nu}}{ 8 \pi \alpha' t}.
\end{align}
\end{subequations}

When $B=0$, in the limit $\alpha' \rightarrow 0$, 
keeping $\Tilde{T} = 2 \pi \alpha' t$ and $\tau_p  \rightarrow \alpha'
\tau_p$ fixed,
the above propagators all reduce to those on a 
circle of length $\Tilde{T}$,
\begin{equation} \label{wlpro}
{\mathcal D}^{\mu\nu}(\tau_p, \tau_q) = - G^{\mu \nu} 
\frac{|\tau_{pq}| ( \Tilde{T}- |\tau_{pq}|)}{\Tilde{T}}.
\end{equation}

However when $\Theta \neq 0$, for nonplanar processes, the worldsheet 
has a finite length $\abs{\Theta k}$ in the $\sigma$-direction, in the limit 
$\alpha' \rightarrow 0$, and there are nontrivial correlations in the 
worldsheet. In this case, we have,
\begin{subequations} \label{fwspro}
\begin{align}
{\mathcal G}^{\mu\nu}(i\tau_i,i\tau_j) 
&= {\mathcal D}^{\mu \nu} (\tau_i, \tau_j)
- \frac{i}{2} \Theta^{\mu \nu} \left[\epsilon (\tau_{ij}) 
       -2 \frac{\tau_{ij}}{\Tilde{T}} \right]; \\ 
{\mathcal G}^{\mu\nu}(\pi + i\tau_x, \pi + i\tau_y) 
& = {\mathcal D}^{\mu \nu} (\tau_x, \tau_y)
+ \frac{i}{2} \Theta^{\mu \nu} \left[ \epsilon (\tau_{xy})
       -2 \frac{\tau_{xy}}{\Tilde{T}} \right]; \\ 
{\mathcal G}^{\mu\nu}(i\tau_i, \pi + i\tau_x) 
& = {\mathcal D}^{\mu \nu} (\tau_i, \tau_x)
+ \frac{(\Theta G \Theta )^{\mu \nu}}{4 \Tilde{T}}.
\end{align}
\end{subequations}

Thus, the field theory limit remains ``stringy''; there is a nontrivial 
two dimensional worldsheet and there are nontrivial
correlations in the worldsheet. After all,  the strong coupling limit of 
this theory is known to be given by a string theory~\cite{gmms,sst} with 
$\apr_{\text{eff}} \sim \Theta$. It is not
inconceivable that one might see stringy effects in perturbation
theory.%
\footnote{S.-J.~Rey (private communication) has pointed out that the
stretched string is reminiscent of the stringy $W$-bosons discussed
in~\cite{br}, and that, in particular, the form of the
$\ast'$-product strongly resembles some stringy structure found
in~\cite{br}.}

\FIGURE{
\includegraphics[height=1in]{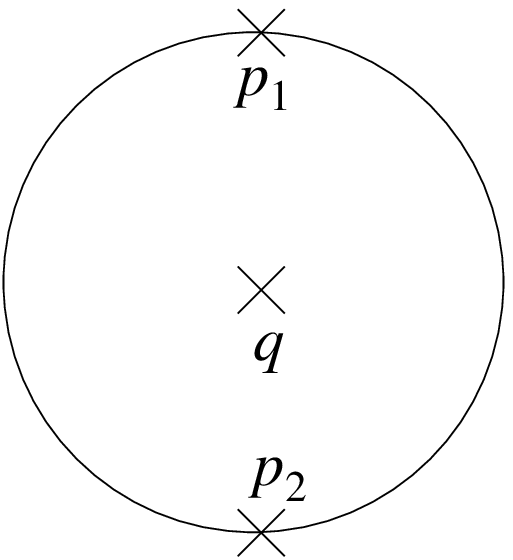}
\caption{The $\ast'$-product also appears in the tree-level
contribution to the interaction of a closed
string with two open strings.\cite{mg} \label{fig:disk}}
}

\skipthis{
Another indication that the theory is stringy is provided by the
following observation.\cite{sjrey} \ The field-theoretic $W$-bosons
that have been
integrated out in writing equation~\eqref{ncf4} are, in the string
theory, fundamental strings stretched between the D-branes.  Since the
string is stretched transverse to the D-branes, the field theory only
sees a point object.  However, it only sees a point object if the
end-points of the string are at the same ($p$+1)-dimensional point;
otherwise one would see two point objects!  Moreover, a disturbance on
one brane should serve to separate the endpoints, for the alternative
requires an acausal instantaneous interaction between the endpoints.
This expectation
has been born out in explicit computation~\cite{br}, and leads to an
improved understanding of causality in the $AdS$/CFT correspondence.
Nevertheless, we do not expect such a stringy effect in ordinary field
theory, because the $W$-boson mass is $r/\apr$, which is
infinitely massive in the field theory limit $\apr\rightarrow 0$,
unless the branes are effectively superposed.  Then, the fundamental
string that is the $W$-boson, is infinitesimally short, and there is
no possible contradiction with causality in assuming that the two
endpoints of the string or at the same  ($p$+1)-dimensional point.  In
the presence of a $B$-field, however, we have already seen that the
mass formulas are modified, and that the stretched string forces us to
insert additional distances into the theory, and thus it may be
natural for the stringy effects observed by Bak and Rey~\cite{br} to
appear even in the field theory limit.  In this case, interestingly,
the string is stretched within the spacetime of the field theory,
rather than transverse!  Furthermore, the differential operation in
the $\ast'$-product bears a
close resemblance to the operator that governs the effect on the
second end-point of the string to a force on the first end-point.%
\footnote{The operator of~\cite{br} is $\frac{\sinh (\Delta x)
\partial}{(\Delta x) \partial} \sim \frac{\sin \lvec[\mu]{\p}
\Theta^{\mu\nu} \rvec[\nu]{\p}}{\lvec[\mu]{\p} \Theta^{\mu\nu}
\rvec[\nu]{\p}}$, which differs from the operator in~\eqref{def*'} by
some factors of 2.}
}

It is well-known that it is a subtle issue to extend 
the first-quantized string theory off-shell. The field variables 
used in string field theory are normally related to those 
in the low energy expansion by complicated field redefinitions.
Here we may have a similar situation. 
It would be interesting  to perform the noncommutative
analog of the computation~\cite{thou} of the one-loop effective action
in field theory, to try to determine, from the field theory point of
view, where off-shell gauge invariance
breaks down.

For a hint at what the appropriate off-shell variables might be which 
respect gauge invariance, we compare our computations to those
in~\cite{mg}, in which the disk diagram in figure~\ref{fig:disk},
describing the interaction of two photons with a massless closed string, was
computed.  Our  one-loop calculation is related 
to the amplitude of figure~\ref{fig:disk} by factorization, so it is
not surprising that in both
cases $\ast'$-products arise. 
Interestingly, Garousi has interpreted the 
$\ast'$-product via the Seiberg-Witten map~\cite{sw}.
(To conform with the notations of~\cite{mg} and~\cite{sw}, we will 
now call our $F$ of previous sections, $\hat{F}$.)

We now briefly summarize the result of Garousi. The disk 
amplitude, figure~\ref{fig:disk}, has an expansion in terms of the
kinematic
invariant $t = -\apr p_1 \cdot p_2$ (the dot is with respect to the 
open string metric). For this amplitude, we can again attempt
to write down a gauge invariant off-shell coupling between the 
massless closed and open string modes, which would reproduce the
leading terms
in $t$. However, a 
closer look at the amplitude reveals a similar problem as in our 1-loop 
case: such a gauge invariant coupling can apparently not be written in
terms of the noncommutative $\Hat{F}$.
\skipthis{
\footnote{In fact, if we could write such a coupling, then we would be
able to find the
NCYM operators corresponding to the modes on the supergravity 
side~\cite{hi,mr}.
(After this work was completed, \cite{dr,ghi} appeared which give the
dual operators in terms of (nonlocal) {\em open\/} Wilson
lines~\cite{iikk,ambjonf,amns2}.  It is
intriguing to note that these Wilson lines are Wilson loops that have
been cut open by precisely the displacement of the stretched string.)} \ 
}
In~\cite{mg},  it was therefore argued
that the appropriate off-shell open string field variables coupling to the 
closed string modes are commutative $F$ and not the noncommutative 
$\hat{F}$ of~\eqref{ncfs}.

It is useful to be more explicit.  Garousi claims that the
lowest order terms in $t$
may be obtained from the following procedure:
\begin{enumerate}
\item Start with the Born-Infeld action with commutative $F$ coupled with
closed string
modes ($h$ below denotes collectively massless closed string modes 
$\phi, b_{\mu \nu}, h_{\mu \nu}$) and expand to second order 
in $F$, {\em i.e.\/} (schematically)
\begin{equation} \label{combi}
{\mathcal L'} \sim h F + h FF + \dots
\end{equation}
\item The Seiberg-Witten map~\cite{sw} gives%
\footnote{Equation~\eqref{swmap} was obtained by integrating 
the infinitesimal form of the Seiberg-Witten transform along a particular 
path.  It is expected that the path dependence\cite{swpath} of the
Seiberg-Witten map can be absorbed into a field redefinition.}
\begin{equation} \label{swmap}
F = \hat{F} + \hat{F} \ast'  \hat{F} + \dots 
\end{equation}
where the above expansion is exact in $\Theta$ and perturbative 
in $\hat{F}$. In the second term $\ast'$ is precisely what we obtained 
in~\eqref{def*'}. 
\item Substitute~\eqref{swmap} into~\eqref{combi} and at the same time 
replace all binary products in~\eqref{combi} between $\Hat{F}$ 
\skipthis{(which is the
only possibility to the second order)}%
by $\ast'$; there is no ambiguity in the order of $\ast'$-products
since only terms up to second order in field strengths are kept.
The 
resulting Lagrangian may be written
\begin{equation} \label{ncbi}
{\mathcal L} \sim h \hat{F} + h \hat{F} \ast' \hat{F} + \dots
\end{equation}  
\item The lowest order terms in $t$ for the diagram in
figure~\ref{fig:disk} (higher
order terms are related to the exchange of massive open string modes) 
are exactly reproduced from the scattering amplitude of $h$ 
and two $\Hat{F}$ calculated from the new Lagrangian ${\mathcal L}$.
\end{enumerate}
Note that the terms in~\eqref{ncbi}, that are quadratic in $\Hat{F}$,
are not gauge invariant, although the original (commutative) DBI
action {\em is\/} gauge invariant.  Hence it is the commutative fields
that are appropriate for describing the process of figure~\ref{fig:disk}.

From the close relation between our one-loop amplitude and the closed string
disk amplitudes, one might wonder whether the commutative $F$ is also 
relevant for  writing down an off-shell gauge invariant 1-loop 
effective action. While this  appears natural from string theory,
it is certainly counterintuitive from the standpoint of noncommutative 
field theory; the field theory hardly knows commutative $F$.
We note that in terms of the commutative field strength, which couples to the
closed string metric, the expansion we were doing in~\eqref{ske} is no 
longer considered a low-energy expansion. Thus it is also not clear 
that we can write down an effective action using $F$ either.
  
Finally, it would be interesting to check whether it is possible to 
recover $\ast'_3$  from the Seiberg-Witten map in the third order term
in~\eqref{swmap}.  It would also be interesting to determine
whether, upon integrating the Seiberg-Witten equation to $n^{\text{th}}$
order in the field strength, one obtains $n$ different
$\ast_m$ $m$-ary operations, $m=1,\dots,n$, or whether this pattern terminates or converges.

\acknowledgments

We have benefited from useful conversation and correspondence with
R.~Britto-Pacumio, C.-S.~Chu, M.~Douglas, M.~Garousi, M.~Li,
J.~Maldacena, G.~Moore, B.~Pioline, A.~Rajaraman, S.-J.~Rey,
M.~Rozali, A.~Sen, A.~Strominger, S.-H. H. Tye and Y.-S. Wu. 
We also thank S.-J.~Rey for a critical reading of a previous draft.
J.M. thanks the Harvard group, where a portion of this work was
performed, for hospitality and fruitful discussion. H.L. thanks 
Center for Advanced Study Tsinghua University, Institute
for Theoretical Physics at Beijing and the string group at Seoul National 
University for hospitality during the last stage of the work.	
This research was supported by DOE grant
\hbox{\#DE-FG02-96ER40559} and an NSERC PDF fellowship.

\appendix

\section{Fermions in a $B$-field} \label{sec:fermions}

The fermionic string in the presence of a $B$-field was discussed 
in~\cite{chuho,scho,sw,cz,ulf}.  
Here we shall give a self-contained discussion of the fermionic
worldsheet fields of the NSR formalism, in the presence of a
(constant) $B$-field and a worldsheet boundary.  For concreteness, we
take the worldsheet to be the strip, parameterized by
\hbox{$w=\sigma+i\tau$};
\hbox{$\sigma\in[0,\pi], \tau\in\field{R}\text{ or }S^1$}.

The bulk action for the fermions $\psi^\mu$ and $\Tilde{\psi}^\mu$ 
is
\begin{equation} \label{spsi}
S = \frac{1}{4\pi \apr} \int d^2w \left\{
(g+B)_{\mu\nu} \psi^\mu \Bar{\p} \psi^\nu
+ (g-B)_{\mu\nu} \Tilde{\psi}^\mu \p\Tilde{\psi}^\nu \right\}.
\end{equation}
The bulk action is
invariant under two supersymmetries, under which, 
\begin{align} \label{2susy}
\delta X^\mu &= -\epsilon \psi^\mu - \Bar{\epsilon} \tilde{\psi}^{\mu} &
\delta \psi^\mu &= \epsilon \p X^{\mu} &
\delta \Tilde{\psi}^\mu &= \Bar{\epsilon} \Bar{\p} X^{\mu},
\end{align}
where $\epsilon$ is the complex supersymmetry parameter.  
When $B=0$, variation of the action gives%
\skipthis{
\footnote{On the strip, this differs by a factor of $i$ from~\cite{jp}.}
}
\begin{equation} \label{psibdy,B=0}
\psi^\mu=\pm  i\Tilde{\psi}^\mu 
\end{equation}
at the boundary; this combined with the
(Neumann) bosonic boundary condition leads to the preservation of only the
supersymmetry for which $\epsilon=\mp i \Bar{\epsilon}$.
We can choose the $+$-sign at one boundary; the choice of sign on the second
boundary then gives us the Neveu-Schwarz (NS) or Ramond (R) sectors, for
respectively the
$-$ or $+$ signs.  In the NS sector, the two boundaries preserve
different supersymmetries, and so globally, no (worldsheet)
supersymmetry is preserved.

For general $B$, the bosonic action
leads to the boundary
condition
\begin{equation} \label{bdyX}
\evalat{g_{\mu\nu}\p_n X^\nu - i B_{\mu\nu}\p_{\tau}X^\nu}{\text{bdy}} 
= 0,
\end{equation}
where $\p_n$ and $\p_\tau$ are respectively normal and tangential
derivatives.
The unbroken supersymmetry---namely that for which
$\epsilon = \mp i \Bar{\epsilon}$---%
and equation~\eqref{bdyX}, leads to
\begin{equation} \label{bdypsi}
(g+B)_{\mu\nu} \psi^\nu=\pm i (g-B)_{\mu\nu} \Tilde{\psi}^\nu,
\end{equation}
which generalizes~\eqref{psibdy,B=0} to $B\neq0$.
We have derived equation~\eqref{bdypsi} from supersymmetry and not
from the action; for a derivation from the action, see~\cite{ulf}.
We can define $\Tilde{\psi}$ so that the $+$-sign holds
at $\sigma=\pi$.  Then we can formally extend the strip to be periodic in
$\sigma$ with periodicity $2\pi$, by setting $\psi^\mu(\sigma,\tau) =
i\bigl(\frac{g-B}{g+B}\bigr)^\mu{_\nu} \Tilde{\psi}^{\nu}(2\pi-\sigma,\tau)$.
The boundary
condition at $\sigma=0$ becomes (anti)periodicity of $\psi$; this is
just the doubling trick extended to $B\neq0$.

To summarize, the only change in the manipulation of the fermions
when $B\neq0$ is in the details of the doubling trick.  The unbroken
supersymmetry is unchanged by the background field, and so the
boundary fields at $\sigma=0,\pi$, $(X^\mu(\tau),\Psi^\mu(\tau))$, which obey
\begin{align}
\delta X^\mu(\tau) &= -\epsilon \Psi^\mu(\tau),
&\delta \Psi^\mu(\tau) = \epsilon \Dot{X}^{\mu}(\tau),
\end{align}
are given by
\begin{equation} \label{defbdypsi}
\begin{split}
\Psi^\mu(0,\tau) 
= \frac{1}{2} [\psi^\mu(0,\tau) \pm i\Tilde{\psi}^\mu(0,\tau)] =
\left(\frac{1}{g-B}\right)^{\mu}{_\nu} \psi^{\nu}(0,\tau)
\\ \Psi^\mu(\pi,\tau) 
= \frac{1}{2} [\psi^\mu(\pi,\tau) + i\Tilde{\psi}^\mu(\pi,\tau)] =
\left(\frac{1}{g-B}\right)^{\mu}{_\nu} \psi^{\nu}(\pi,\tau)
\end{split}
\end{equation}
where, on the rightmost-side of~\eqref{defbdypsi}, $\psi$ is given by
that after
the doubling trick.

The partition function for the various spin structures is given
by~\cite{jp}
\begin{subequations} \label{ppsi}
\begin{align} 
\label{ppsi+-}
\langle 1 \rangle_{+-} &= \frac{\vartheta_2^4(0|it)}{\eta(it)^4}, &
\langle 1 \rangle_{--} &= \frac{\vartheta_3^4(0|it)}{\eta(it)^4}, \\
\label{ppsi-+}
\langle 1 \rangle_{-+} &= \frac{\vartheta_4^4(0|it)}{\eta(it)^4}, &
\langle 1 \rangle_{++} &= \frac{\vartheta_1^4(0|it)}{\eta(it)^4},
\end{align}
\end{subequations}
where the subscript ${\pm\pm'}$ denotes the periodicity in
the $\sigma$ and $\tau$ directions respectively. 
The R (NS) sector
is (anti)periodic in $\sigma$ and antiperiodic in $\tau\sim\tau+2\pi
t$; the corresponding $(-1)^F$
sectors are periodic in $\tau$.

The Green
functions in the NS sector for the fermions 
on the annulus
are determined, via the
doubling trick, by
holomorphy, (anti)periodicity and the OPE 
\hbox{$\psi^\mu (w) \psi^\nu (w')\sim \frac{\apr}{2}
\frac{g^{\mu\nu}}{w-w'}$}.
The Green functions are
thus~\cite{nns}
\begin{subequations} \label{psicorr}
\begin{align}
\label{psicorr+-}
\vev{\psi^\mu(w_1) \psi^{\nu}(w_2)}_{+-} 
&= \frac{\apr}{4\pi} g^{\mu\nu}
   \frac{\vartheta_2\bigl(\frac{w_1-w_2}{2\pi}|it\bigr)\vartheta_1'(0|it)}%
        {\vartheta_1\bigl(\frac{w_1-w_2}{2\pi}|it\bigr)\vartheta_2(0|it)},
\\
\label{psicorr--}
\vev{\psi^\mu(w_1) \psi^{\nu}(w_2)}_{--} 
&= \frac{\apr}{4\pi} g^{\mu\nu}
   \frac{\vartheta_3\bigl(\frac{w_1-w_2}{2\pi}|it\bigr)\vartheta_1'(0|it)}%
        {\vartheta_1\bigl(\frac{w_1-w_2}{2\pi}|it\bigr)\vartheta_3(0|it)},
\\
\label{psicorr-+}
\vev{\psi^\mu(w_1) \psi^{\nu}(w_2)}_{-+} 
&= \frac{\apr}{4\pi} g^{\mu\nu}
   \frac{\vartheta_4\bigl(\frac{w_1-w_2}{2\pi}|it\bigr)\vartheta_1'(0|it)}%
        {\vartheta_1\bigl(\frac{w_1-w_2}{2\pi}|it\bigr)\vartheta_4(0|it)}.
\end{align}
Equation~\eqref{defbdypsi} immediately implies that correlation
functions of the boundary fermions $\Psi$ are given
by~\eqref{psicorr}, with the closed string metric $g^{\mu\nu}$
replaced with the open 
string metric
\hbox{$G^{\mu\nu}=\bigl(\frac{1}{g+B}g\frac{1}{g-B}\bigr)^{\mu\nu}$}.

For general periodicity \hbox{$\psi^\mu(w+2n\pi+2m\pi i t) = e^{2\pi i [n
(\alpha-1/2) + m(\beta-1/2)]}\psi^\mu(w)$}, we can
write~\cite{nns}
\begin{equation} \tag{\ref{psicorr}$'$}
\vev{\psi^\mu(w_1) \psi^{\nu}(w_2)}
= \frac{\apr}{4\pi} g^{\mu\nu}
  \frac{\vartheta\left[\begin{smallmatrix} \alpha \\ \beta
                       \end{smallmatrix} \right]
               \bigl(\frac{w_1-w_2}{2\pi}|it\bigr)
         \vartheta'\left[\begin{smallmatrix} 1/2 \\ 1/2
                       \end{smallmatrix}\right](0|it)}%
        {\vartheta\left[\begin{smallmatrix} 1/2 \\ 1/2 
                       \end{smallmatrix} \right]
               \bigl(\frac{w_1-w_2}{2\pi}|it\bigr)
         \vartheta\left[\begin{smallmatrix} \alpha \\ \beta 
                        \end{smallmatrix}\right](0|it)},
\end{equation}
\end{subequations}
and the reader who is not confused by the observation that
the doubly periodic case of $\alpha=\beta=1/2$ in 
equation~(\ref{psicorr}$'$) is everywhere singular, can skip the rest
of this paragraph.
Of course, for this spin structure, we
cannot write down a Green
function with the (allegedly) desired properties, because of the
theorem~\cite[p.~431]{ww}
that there are no doubly periodic meromorphic functions with a single
simple pole.  Physically, this is the result of the
existence of a
(constant) zero mode on the doubly periodic torus.  Thus
\skipthis{
Wick's
theorem no longer holds in the most straightforward way, and more
importantly for our purposes, 
}%
the doubly periodic spin structure does
not contribute to correlation functions with fewer than 10 fermions.
Since the correlation functions that we consider have no more than 8
fermions, we can simply ignore the doubly periodic spin structure in
this paper.
\skipthis{
Finally, we note that if we evaluate, using the doubling trick, the
Green function on the
boundary, for the boundary fermions $\Psi^\mu$, that the result is
just~\eqref{psicorr}, but replacing the closed string metric
$g^{\mu\nu}$ with the open string metric 
\hbox{$G^{\mu\nu}=\bigl(\frac{1}{g+B}g\frac{1}{g-B}\bigr)%
\skipthis{\left(\frac{1}{g+B}\right)_S}%
^{\mu\nu}$}.
}

\section{A Note on the Boundary Green Function} \label{sec:prop}

In this section we give a derivation of the boundary propagators on the 
annulus in the presence of a $B$-field, since there is  some
confusion in the literature as to what the correct Green function is.
Our discussion will be based on the results in~\cite{cn}.

\skipthis{
It has been emphasized in {\em e.g.\/}~\cite{bcr} that because the
vertex operator involves $\dot{X}^\mu$, the boundary Green function
${\mathcal G}^{\mu\nu}(w_1,w_2)$
does not have the usual ambiguity by an arbitrary function of $w_1$
plus an arbitrary function of $w_2$.  Nevertheless, there is some
confusion in the literature as to what the correct Green function is.
Although we did not need it in either this paper or in~\cite{lm}---the
$\dot{X}^\mu$ term did not survive the computation---we wish to use
this appendix to put in our two cents.  Our conclusion, based on an
analysis of periodicity and branch cuts, is that the
Green function of~\cite{cn} comes the closest, of the results
heretofore given in the literature, to being correct.
}

The goal is to find a solution of
\begin{subequations} \label{propgreen}
\begin{gather}
\label{boxgreen}
\partial_1 \Bar{\partial}_1 \g^{\mu\nu}(w_1,w_2)
= -\pi \alpha' g^{\mu\nu} \delta^{(2)}(w_1-w_2) 
+ \frac{\apr}{4 \pi t} g^{\mu\nu}, \\
\label{bdygreen}
\evalat{\left\{
(g+B)_{\mu \lambda} \partial_1 \g^{\lambda\nu}(w_1,w_2)
+(g-B)_{\mu \lambda} \Bar{\partial}_1 \g^{\lambda\nu}(w_1,w_2) 
\right\}}{\sigma_1=0,\pi} = 0, \\
\label{periodgreen}
\g^{\mu\nu}(w_1+2\pi i t,w_2) = \g^{\mu\nu}(w_1,w_2) =
\g^{\mu\nu}(w_1,w_2+2\pi i t).
\end{gather}
\end{subequations}
Equation~\eqref{boxgreen} is the equation of motion
for $X^\mu$ and eq.~\eqref{bdygreen} is the boundary condition for
$X^\mu(w_1)$, as modified by the $B$-field in the now-familiar way.
Symmetry, $\g^{\mu\nu}(w_1,w_2)=\g^{\nu\mu}(w_2,w_1)$ 
enforces the boundary condition on $X^\nu(w_2)$, so we do not need to
impose that separately.
The last term in equation~\eqref{boxgreen} is a background charge
without which~\eqref{boxgreen} would be inconsistent
with~\eqref{bdygreen} and Gauss' law~\cite{jp}.  
Finally, the periodicity condition~\eqref{periodgreen} enforces
singlevaluedness of the Green function on the annulus.

Chaudhury and Novac~\cite{cn} solve equations~\eqref{propgreen} with%
\footnote{A similar formula in ref.~\cite{shen} does not have the
linear
terms in each of the second and third lines.  These terms are related
by the boundary condition~\eqref{bdygreen} and do not affect the
equation of motion~\eqref{boxgreen}.  Our careful analysis of
periodicity~\eqref{periodgreen}
shows that these terms are essential.\label{ft:shen}}
\begin{multline} \label{green}
\g^{\mu\nu}(w_1,w_2) =
-\frac{\alpha'}{2} G^{\mu\nu} \left\{ 
\ln \abs{\frac{\vartheta_1\bigl(\frac{w_1-w_2}{2\pi it}|\frac{i}{t}\bigr)
\vartheta_1\bigl(\frac{w_1+\bar{w}_2}{2\pi it}|\frac{i}{t}\bigr)}%
{\eta^2\bigl(\frac{i}{t}\bigr)}}^2
  - \frac{(\real w_1)^2 + (\real w_2)^2}{\pi t} \right\} \\
- \frac{\Theta^{\mu\nu}}{2\pi} \left\{
\log \vartheta_1\bigl(\frac{w_1+\bar{w}_2}{2\pi i t}|
   \frac{i}{t}\bigr)
-\log \vartheta_1\bigl(\frac{\bar{w}_1+w_2}{2\pi i
 t}|\frac{i}{t}\bigr)
-\frac{i}{t} \im(w_1-w_2)
\right\} \\
+ \frac{(\Theta G \Theta)^{\mu\nu}}{8\pi^2\apr} \left\{
\ln \abs{\frac{\vartheta_1\bigl(\frac{w_1-w_2}{2\pi it}|\frac{i}{t}\bigr)}%
{\vartheta_1\bigl(\frac{w_1+\bar{w}_2}{2\pi it}|\frac{i}{t}\bigr)}}^2
- \frac{[\real w_1 - \pi]^2 + [\real w_2 - \pi]^2}{\pi t} + \frac{2
\pi}{t} \right\}.
\end{multline}
However, equation~\eqref{green} is not complete without specifying a branch 
for the logarithm for which the Green function is
continuous. In equation~\eqref{green} the dangerous logarithms
reside in the function multiplying $\Theta^{\mu \nu}$. As the arguments 
of the logarithms cross the branch cut, there are possible discontinuities
in the Green function, which will result in  
extra, $\delta'$-function sources on the right hand side of
equation~\eqref{boxgreen}.
Thus, as it is written here, the Green function does not
actually obey the equation of motion~\eqref{boxgreen}.
 
It would be  desirable to choose a branch cut that is never crossed in 
the range of the variables of the Green function; then the
Green function would automatically be continuous. 
An example of this is the 
Green function for the upper-half plane
with the branch cut along the positive real 
axis.%
\footnote{The boundary is not considered part of the worldsheet; in
particular, we define the boundary Green function by taking the limit
from the interior.}
However, for the annulus, the nature of the $\theta$-functions is such that
there is no choice of branch cut that is never crossed. 
Thus the function multiplying $\Theta^{\mu \nu}$ in the second 
line of equation~\eqref{green} is discontinuous, and will produce
additional $\delta'$-functions proportional to $\Theta^{\mu\nu}$ on the
right hand side of the equation of motion~\eqref{boxgreen}. The remedy is to
add a function to equation~\eqref{green} to explicitly cancel the 
unphysical discontinuities (of course, we will ensure that this function 
affects neither the equation of motion nor the boundary condition).  

\skipthis{
Recall that on the disk, the
$\Theta$-dependence of the boundary Green function arises solely from
a branch cut analysis.\cite{sw} \  Explicitly, the bulk Green function
on the upper half-plane ($\uhp$) is
\begin{equation} \label{diskgreen}
\begin{split}
\g_{\uhp}^{\mu\nu}(z_1,z_2) = 
- \apr &\left[ g^{\mu \nu} (\log\abs{z_1-z_2} - \log\abs{z_1- \bar{z}_2})
+ G^{\mu \nu} \log \abs{z_1 - \bar{z}_2}^2 
\right. \\ & \left.
+ \frac{1}{2 \pi \apr} \Theta^{\mu \nu} 
   \left( \pi i+\log {\frac{z_1 - \bar{z}_2}{\bar{z}_1-z_2}}\right)
\right].
\end{split}
\end{equation}
This is not well-defined without a choice a branch for the logarithm;
we choose to put the branch cut along the positive real axis, such that
$\im \log z \in [0,2\pi)$.
We will denote this by writing
$\Log_{[0,2\pi)}z$.
This is the unique choice which ensures that for $z_{1,2}=y_{1,2}+i
x_{1,2}\in\uhp$, the
branch cut
is never crossed in~\eqref{diskgreen}, where it is understood that
the real axis is not part of $\uhp$.%
\skipthis{
\footnote{More explicitly, note that if $z=y_1+i x_1$ and $w=y_2+i
x_2$ then ${\frac{z - \bar{w}}{\bar{z}-w}} =
\frac{(y_1-y_2)^2-(x_1+x_2)^2+2i
(y_1-y_2)(x_1+x_2)}{\abs{x-\bar{w}}^2}$,
which has arbitrary phase, except that it is never on the positive
real axis when $x_1,x_2>0$.\label{ft:expllog}}
}
Then, as $z_1$ and $z_2$ approach the real axis from
the bulk of
$\uhp$---that is, $x_{1,2} \approx \delta\rightarrow 0^+$---%
equation~\eqref{diskgreen} reads
\begin{smaleq}
\begin{equation} \label{diskbdygreen}
\begin{split}
\g_{\uhp}^{\mu\nu}(y_1,y_2) &= 
- \apr G^{\mu\nu} \log \abs{y_1-y_2}^2
- \frac{1}{2\pi\apr} \Theta^{\mu\nu} \left[\Log_{[0,2\pi)} 
 \left(1+i \frac{\delta}{y_1-y_2}+ \order{\delta^2}\right)
 - \pi i \right] \\
&= - \apr G^{\mu\nu} \log \abs{y_1-y_2}^2
+ i\frac{1}{2\apr} \Theta^{\mu\nu} \epsilon(y_1-y_2).
\end{split}
\end{equation}
\end{smaleq}%
Had the branch cut been neglected, noncommutativity would have
remained unobserved.
}

The simplest choice of branch cut appears to be along the negative real
axis.
We choose $-\pi<\im \log z \leq \pi$ and denote this by 
writing $\Log_{(-\pi ,\pi]}$. 
It can be seen from the expression
\begin{equation} \label{sumthetareim}
\begin{split} 
\vartheta_1\bigl(\frac{\sigma+i \tau}{2\pi i t}|\frac{i}{t}\bigr)
 &= 2 \sum_{n=0}^{\infty} (-1)^n e^{-\frac{\pi}{t} (n+\frac{1}{2})^2}
   \left[\sin\bigl((n+\frac{1}{2})\frac{\tau}{t}\bigr)
         \cosh\bigl((n+\frac{1}{2})\frac{\sigma}{t}\bigr) 
\right. \\ & \qquad \qquad \left.
       -i\cos\bigl((n+\frac{1}{2})\frac{\tau}{t}\bigr)
         \sinh\bigl((n+\frac{1}{2})\frac{\sigma}{t}\bigr)
   \right],
\end{split}
\end{equation}
that this branch cut is crossed, in a counterclockwise direction, by
$\vartheta_1\bigl(\frac{w_1+\bar{w}_2}{2\pi i t}|\frac{i}{t}\bigr)$ when
$\tau_{12}$ increases through $\tau_{12}=(4n-1)\pi t$ for $n \in {\mathbb{Z}}$.
(Recall our notation is $w_i = \sigma_i+i \tau_i$ where $\sigma_i\in(0,\pi)$
and $\tau_i \in {\mathbb{R}}$; thus $0<\sigma_1+\sigma_2<2\pi$.)
Similarly, $\vartheta_1\bigl(\frac{\bar{w}_1+w_2}{2\pi i
t}|\frac{i}{t}\bigr)$ crosses the branch cut in a clockwise direction
as $\tau_{12}$ increases through $\tau_{12}=(4n+1)\pi t$.
We believe that these are the only crossings.
Thus the discontinuities of the function multiplying 
$\Theta^{\mu \nu}$ in the second line of equation~\eqref{green},
\begin{equation} \label{fef}
\Log_{(-\pi,\pi]} \vartheta_1\bigl(\frac{\sigma+i\tau}{2\pi i t}\bigr)
-\Log_{(-\pi,\pi]} \vartheta_1\bigl(\frac{\sigma-i\tau}{2\pi i
 t}\bigr),
\end{equation}
can be mimicked by a function $2 \pi i f(\tau)$ with $f$ defined by
\begin{equation} \label{deff}
f(\tau) = - \left[\frac{\tau}{2 \pi t}\right],
\end{equation}
where $[x]$ denotes the closest integer to $x$.%
\footnote{%
Continuity of equation~\eqref{usegreen} further sets
$[\frac{1}{2}+2n]=2n$ and
$[\frac{3}{2}+2n]=2(n+1)$ for $n\in {\mathbb{Z}}$.\label{ft:f[1/2]}}
\skipthis{
This arises as follows: recall from equation~\eqref{sumthetareim} that the
first term of equation~\eqref{fef} crosses the branch
cut for $\tau = (4 n -1)\pi t$.  As $\tau$ increases across the branch
cut, $\Log_{(-\pi,\pi]} \vartheta_1\bigl(\frac{\sigma+i\tau}{2\pi i
t}\bigr)$ decreases by $2\pi i$.  The otherwise constant
function $-2\pi i\left[\frac{\tau}{2\pi t}\right]$ also decreases by $2\pi
i$ at the half-integer $\tau=(4n-1)\pi t$.  The similar decrease by
$2\pi i$ of $2\pi i f(\tau)$ at $\tau=(4n+1)\pi t$ is the same as the
jump in the second term of equation~\eqref{fef}, as follows by
reversing all the signs (and switching ``increase'' and ``decrease'')
in the previous two sentences.}

Thus, the difference of equation~\eqref{fef} and $2\pi i f(\tau)$ is
continuous across the branch cut, but since $2\pi i f(\tau)$ is
otherwise constant, it preserves the (differential) properties of
equation~\eqref{fef} away from the branch cut.
So, by replacing equation~\eqref{green} with
\begin{multline} \label{usegreen}
\g^{\mu\nu}(w_1,w_2) =
-\frac{\alpha'}{2} G^{\mu\nu} \left\{ 
\ln \abs{\frac{\vartheta_1\bigl(\frac{w_1-w_2}{2\pi it}|\frac{i}{t}\bigr)
\vartheta_1\bigl(\frac{w_1+\bar{w}_2}{2\pi it}|\frac{i}{t}\bigr)}%
{\eta^2\bigl(\frac{i}{t}\bigr)}}^2
  - \frac{(\real w_1)^2 + (\real w_2)^2}{\pi t} \right\} \\
- \frac{\Theta^{\mu\nu}}{2\pi} \left\{
\Log_{(-\pi,\pi]} \vartheta_1\bigl(\frac{w_1+\bar{w}_2}{2\pi i t}|
   \frac{i}{t}\bigr)
-\Log_{(-\pi,\pi]} \vartheta_1\bigl(\frac{\bar{w}_1+w_2}{2\pi i
 t}|\frac{i}{t}\bigr) 
      \right. \\ \left.
-\frac{i}{t} \im(w_1-w_2)
- 2\pi i f(\im(w_1-w_2))
\right\} \\
+ \frac{(\Theta G \Theta)^{\mu\nu}}{8\pi^2\apr} \left\{
\ln \abs{\frac{\vartheta_1\bigl(\frac{w_1-w_2}{2\pi it}|\frac{i}{t}\bigr)}%
{\vartheta_1\bigl(\frac{w_1+\bar{w}_2}{2\pi it}|\frac{i}{t}\bigr)}}^2
- \frac{[\real w_1 - \pi]^2 + [\real w_2 - \pi]^2}{\pi t} + \frac{2
\pi}{t} \right\},
\end{multline}
we now satisfy equations~\eqref{propgreen}, even across the branch cut.

Having modified the antisymmetric part of~\eqref{green} by some step
functions, we should check that it indeed obeys the periodicity 
condition~\eqref{periodgreen}.  (The
periodicity of the symmetric part of~\eqref{usegreen} is obvious.)
It is enough to check periodicity under $w_1\rightarrow w_1+2\pi i t$,
since the Green function depends on the imaginary parts of $w_1$ and
$w_2$ only as $\im(w_1-w_2)$; this is translational invariance along
the cylinder.  Under $w_1\rightarrow w_1+2\pi t$, the
$\vartheta_1$-function changes sign; this is the standard
antiperiodicity of $\vartheta_1$.  If, for some $n\in{\mathbb{Z}}$,
\hbox{$(4n-1)\pi t < \im(w_1-w_2)<(4n+1)\pi t$}, then the first $\Log$
gives an additional $\pi i$ (the translation of $w_1$
yields a counterclockwise
rotation of the $\vartheta$-function which does not cross the branch
cut). The second $\Log$ is essentially
a reflection across the imaginary axis, and thus is a clockwise
rotation that does cross the branch cut; this then also gives $\pi i$
and the difference vanishes.  Similarly, the difference of the
$\Log \vartheta_1$s also cancels in the complementary region
$(4n+1)\pi t < \im(w_1-w_2) <
(4n+3)\pi t$.%
\footnote{This leaves $\im(w_1-w_2) = (2 n+1) \pi t$.  This is more
involved and the specification of footnote~\ref{ft:f[1/2]} is vital.}
Finally, it is clear that $2\pi if(\im(w_1-w_2))$ jumps by $-2\pi i$ under a
shift of $w_1$ by $2 \pi t$,%
\footnote{Again the analysis is more involved when
$\im(w_1-w_2) = (2 n+1) \pi t$.}
which cancels the shift in $\im(w_1-w_2)$.  Thus the Green
function~\eqref{usegreen} is periodic.  Note that the explicit
inclusion of $\im(w_1-w_2)$ was crucial (cf.\ footnote~\ref{ft:shen}).

\skipthis{
\FIGURE[t]{
\includegraphics[width=3.25in,height=2in]{imbranch.eps}
\caption{The location of ${\frac{z_1 - \bar{z}_2}{\bar{z}_1-z_2}} =
\frac{(y_1-y_2)^2-(x_1+x_2)^2+2i
(y_1-y_2)(x_1+x_2)}{\abs{z_1-\bar{z}_2}^2}$. \label{fig:imbranch}}
}

Since the branch cut is crossed, equation~\eqref{green} is actually
discontinuous.  Again it is worthwhile illustrating our point with the
more familiar
disk.  Suppose we had chosen our branch cut on the negative imaginary
axis.  Then, the
branch is crossed at $(y_2-y_1)=(x_1+x_2)$ (see
figure~\ref{fig:imbranch}).
At this arbitrary location, the value of the logarithm jumps by $2\pi i$.
We must explicitly cancel this unphysical
discontinuity in the Green function by writing
\begin{multline} \label{moddiskgreen}
\g_{\uhp}^{\mu\nu}(z_1,z_2) = 
- \apr \Bigl[ g^{\mu \nu} (\log\abs{z_1-z_2} - \log\abs{z_1- \bar{z}_2})
+ G^{\mu \nu} \log \abs{z_1 - \bar{z}_2}^2 
\\
+ \frac{1}{2 \pi \apr} \Theta^{\mu \nu} 
   \left( \pi i+
   \Log_{[-\frac{\pi}{2},\frac{3\pi}{2})} {\frac{z_1 -
   \bar{z}_2}{\bar{z}_1-z_2}}
   - 2 \pi i \theta\bigl(\im(z_1-\Bar{z}_2)+\real(z_1-\Bar{z}_2)\bigr) \right)
\Bigr],
\end{multline}
where $\theta(z)$ is a%
\footnote{Continuity further sets $\theta(0)$=0.\label{ft:step0}}
unit step function with opposite discontinuity from that in the
logarithm so that
the Green function~\eqref{moddiskgreen} is continuous across the
branch cut.  Since the step function is otherwise constant, it does
not contribute anything else to the equation of motion or the boundary
condition.
Upon taking $z_1$ and $z_2$ to the boundary, the logarithm is
continuous, but the value of the step function now depends on whether
$y_1-y_2$ is positive or negative.  This reproduces
equation~\eqref{diskbdygreen}.
}

Equations~\eqref{wspro} then follow upon taking the limit
that $w_1$, $w_2$ are on the boundary, in the region $0\leq
\tau_1,\tau_2<2\pi t$.  Outside this region, the actual expression is more
complicated, but it follows from periodicity.

\begin{description}
\item{\protect $\mathbf{\sigma_1=\sigma_2=0}$:}
First consider $-\pi t <
\tau_{12} <\pi t$.  In this region, one can show that
\begin{equation} \label{t00}
\vartheta_1 
\bigl(\frac{\delta + i \tau_{12}}{2\pi i t}|\frac{i}{t}\bigr)
\sim - i\delta + \epsilon (\tau_{12}).
\end{equation}
\skipthis{
Thus,
\begin{subequations}
\begin{align}
\Log \left[ \vartheta_1 
\bigl(\frac{\delta + i \tau_{12}}{2\pi i t}|\frac{i}{t}\bigr) \right]
&=  \begin{cases}   0 , &  \tau_{12} > 0 \\
                - i \pi  & \tau_{12} < 0 
                         \end{cases},
\quad \abs{\tau_{12}}<\pi t, \\
\Log \left[ \vartheta_1 
\bigl(\frac{\delta - i \tau_{12}}{2\pi i t}|\frac{i}{t}\bigr) \right]
&=  \begin{cases}  -i \pi , &  \tau_{12} > 0 \\
                0  & \tau_{12} < 0
                         \end{cases},
\quad \abs{\tau_{12}}<\pi t.
\end{align}
\end{subequations}
}
Therefore,
\begin{equation} \label{pg11}
{\mathcal G}^{\mu\nu}(i\tau_1,i\tau_2) = -\alpha' G^{\mu\nu} \ln
\abs{\frac{\vartheta_1\bigl(\frac{\tau_{12}}{2\pi t}|\frac{i}{t}\bigr)}%
{\eta\bigl(\frac{i}{t}\bigr)}}^2
+ \frac{i}{2} \Theta^{\mu\nu} \left [ \frac{\tau_{12}}{\pi t}
-  \epsilon(\tau_{12}) \right] , \;\;
-\pi t<\tau_{12} < \pi t,
\end{equation}
Periodicity extends this result to $-2\pi t<\tau_{12}<2\pi t$;
remarkably, the form is the same.  (However, outside this larger
region, the form changes.)  This reproduces equation~\eqref{wspro00}.%
\footnote{We have dropped an overall $2 \apr G^{\mu\nu} \log
\eta\bigl(\frac{i}{t}\bigr)$ from each of equations~\eqref{wspro};
this is just a ($t$-dependent) constant.}

\item{\protect$\mathbf{\sigma_1=\sigma_2=\pi}$:} Quasi-periodicity of
the $\vartheta$-function (see e.g.~\cite{ww}) gives
\begin{equation} \label{appeo}
\vartheta_1\bigl(\frac{\tau_{12}+ i \delta}{2\pi t} 
- \frac{i}{t}|\frac{i}{t}\bigr)
 =  - e^{\frac{\pi}{t} - \frac{\delta}{t} + i \frac{\tau_{12}}{t}}
\vartheta_1\bigl(\frac{\tau_{12}+ i \delta}{2\pi t}|\frac{i}{t}\bigr).
\end{equation}
Thus, for $-\pi t<\tau_{12}<\pi t$, and using equation~\eqref{t00},
\begin{equation}
\begin{split}
\Log_{(-\pi,\pi]} 
  \vartheta_1\bigl(\frac{2\pi -\delta+i\tau_{12}}{2\pi t}|\frac{i}{t}\bigr)
&- \Log_{(-\pi,\pi]} 
  \vartheta_1\bigl(\frac{2\pi -\delta-i\tau_{12}}{2\pi
t}|\frac{i}{t}\bigr)
\\
\skipthis{
&= \Log_{(-\pi,\pi]} -e^{\frac{\pi}{t} - \delta + i \frac{\tau_{12}}{t}} 
         (-i \delta + \epsilon(\tau_{12})
- \Log_{(-\pi,\pi]} -e^{\frac{\pi}{t} - \delta - i \frac{\tau_{12}}{t}} 
         (-i \delta - \epsilon(\tau_{12}) \\}
&= -\pi i \epsilon(\tau_{12}) + 2 i  \frac{\tau_{12}}{t}.
\end{split}
\end{equation}
Hence,
\begin{equation} \label{pg22}
\!\!\!\!\!\!\!\!
{\mathcal G}^{\mu\nu}(\pi+i\tau_1,\pi+i\tau_2) = -\alpha' G^{\mu\nu} \ln
\abs{\frac{\vartheta_1\bigl(\frac{\tau_{12}}{2\pi t}|\frac{i}{t}\bigr)}%
{\eta\bigl(\frac{i}{t}\bigr)}}^2
- \frac{i}{2} \Theta^{\mu\nu} \left [ \frac{\tau_{12}}{\pi t}
-  \epsilon(\tau_{12}) \right] , \;\;
-\pi t<\tau_{12} < \pi t;
\end{equation}
again, this has the same form when extended to $-2\pi t<\tau_{12}<2\pi
t$.  So we have equation~\eqref{wspro11}.

\item{\protect$\mathbf{\sigma_1=0, \sigma_2=\pi}$:}
Using
\begin{equation}
\vartheta_1\bigl(\frac{w_1 +\bar{w}_2}{2\pi i t}|
   \frac{i}{t}\bigr) 
= \vartheta_1\bigl(\frac{\tau_{12} - i \delta}{2\pi t} 
- \frac{i}{2t}|\frac{i}{t}\bigr)
= - i e^{\frac{\pi}{4t} +  i \frac{\tau_{12}}{2t}}
\vartheta_4\bigl(\frac{\tau_{12}- i \delta}{2\pi t}|\frac{i}{t}\bigr)
\end{equation}
and the fact that
$\vartheta_4\bigl(\frac{\tau_{12}}{2\pi t}|\frac{i}{t}\bigr)>0$
for all (real) $\tau_{12}$, we immediately find
\begin{equation} \label{pg12}
{\mathcal G}^{\mu\nu}(i\tau_1,\pi + i\tau_2) = -\alpha' G^{\mu\nu} \ln
\abs{\frac{\vartheta_4\bigl(\frac{\tau_{12}}{2\pi t}|\frac{i}{t}\bigr)}%
{\eta\bigl(\frac{i}{t}\bigr)}}^2 +  
\frac{(\Theta G \Theta)^{\mu\nu}}{8\pi \apr t}.
\end{equation}
In particular, the antisymmetric part vanishes, so periodicity is trivial.
This is equation~\eqref{wspro01}.
Note that $\sigma_1=\pi,\sigma_2=0$ is exactly the same
since $\real(w_1+\bar{w}_2)=\sigma_1+\sigma_2$.

\end{description}

Finally, one can check,
though one must be careful to keep certain self-contractions, that the
two-point function for two gauge bosons in the bosonic string
correctly reproduces the field theory one-loop contribution to the
two-point function.  This was the evidence that caused~\cite{bcr} to
assert the correctness of their boundary Green function, even though that one
turns out to not be periodic.

\end{document}